\documentclass[a4paper,fleqn,usenatbib]{mnras}

\usepackage{newtxtext,newtxmath}
\usepackage[T1]{fontenc}
\usepackage{ae,aecompl}
\usepackage{url}
\usepackage{graphicx}	% Including figure files
\usepackage{amsmath}	% Advanced maths commands
\usepackage{amssymb}	% Extra maths symbols

\title[Cyg X-3 during the 2016 giant flare]{Single-dish and VLBI observations of Cygnus X-3 during the 2016 giant flare episode}

\author[Egron et al.]{
E. Egron$^{1}$\thanks{E-mail: egron@oa-cagliari.inaf.it},
A. Pellizzoni$^{1}$,
M. Giroletti$^{2}$,
S. Righini$^{2}$,
M. Stagni$^{2}$,
A. Orlati$^{2}$,
\newauthor C. Migoni$^{1}$,
A. Melis$^{1}$,
R. Concu$^{1}$,
L. Barbas$^{3}$, %(Yebes)
S. Buttaccio$^{4}$, % (Noto) 
P. Cassaro$^{4}$, %(Noto)
\newauthor P. De Vicente$^{3}$, % (Yebes)
M.P. Gawro\'{n}ski$^{5}$, %(Torun)
M. Lindqvist$^{6}$, %(Onsala)
G. Maccaferri$^{2}$,
C. Stanghellini$^{2}$, %(Noto)
\newauthor P. Wolak$^{5}$, % (Torun) 
J. Yang$^{6}$, %(Osala)
A. Navarrini$^{1}$,
S. Loru$^{1}$,
M. Pilia$^{1}$,
M. Bachetti$^{1}$,
M.N. Iacolina$^{7,1}$,
\newauthor M. Buttu$^{1}$,
S. Corbel$^{8,9}$,
J. Rodriguez$^{8}$, 
S. Markoff$^{10}$,
J. Wilms$^{11}$,
K. Pottschmidt$^{12,13}$,
\newauthor M. Cadolle Bel$^{14}$,
E. Kalemci$^{15}$,
T. Belloni$^{16}$,
V. Grinberg$^{17}$,
M. Marongiu$^{18,1}$,
\newauthor G.P. Vargiu$^{1}$,
A. Trois$^{1}$\\
$^{1}$INAF, Osservatorio Astronomico di Cagliari, Via della Scienza 5, 09047 Selargius, Italy\\
$^{2}$INAF, Istituto di Radio Astronomia di Bologna, Via P. Gobetti 101, 40129 Bologna, Italy\\
$^{3}$Centro Nacional de Tecnolog\'{i}as Radioastron\'{o}micas y Aplicaciones Geoespaciales(CNTRAG), Observatorio de Yebes (IGN), Spain\\
$^{4}$INAF, Istituto di Radioastronomia, Sezione di Noto, Contrada Renna Bassa, 96017 Noto, Italy\\
$^{5}$Toru\'{n} Centre for Astronomy, N. Copernicus University, Gagarina 11, 87-100 Toru\'{n}, Poland\\
$^{6}$Department of Earth and Space Sciences, Chalmers University of Technology, Onsala
Space Observatory, 439 92 Onsala, Sweden\\
$^{7}$Agenzia Spaziale Italiana - Via del Politecnico snc 00133 Roma, Italy\\
$^{8}$Laboratoire AIM, UMR 7158, CEA/CNRS/Universit\'{e} Paris Diderot, CEA DRF/IRFU/DAp, 91191 Gif-sur-Yvette, France\\
$^{9}$Station de Radioastronomie de Nan\c{c}ay, Observatoire de Paris, PSL Research University, CNRS, Univ. Orl\'{e}ans, 18330 Nan\c{c}ay, France\\
$^{10}$Anton Pannekoek Institute for Astronomy, University of Amsterdam, PO Box 94249, 1090 GE Amsterdam, The Netherlands\\
$^{11}$Dr.\,Karl-Remeis-Sternwarte and Erlangen Centre for Astroparticle Physics (ECAP), Friedrich Alexander Universit\"{a}t Erlangen-N\"{u}rnberg, \\Sternwartstr. 7, 96049 Bamberg, Germany\\
$^{12}$CRESST and NASA Goddard Space Flight Center, Astrophysics Science Division, Code 661,
Greenbelt, MD 20771, USA\\
$^{13}$Center for Space Science and Technology, University of Maryland Baltimore County, 1000
Hilltop Circle, Baltimore, MD 21250, USA\\
$^{14}$Max Planck Computing and Data Facility, 85748 Garching, Germany\\
$^{15}$Faculty of Engineering and Natural Sciences, Sabanc\i\ University, Orhanl\i\ -Tuzla, 34956 Istanbul, Turkey\\
$^{16}$INAF, Osservatorio Astronomico di Brera, via E. Bianchi 46, 23807 Merate, Italy \\
$^{17}$ESA/ESTEC, Keplerlaan 1, 2201 AZ Noordwijk, The Netherlands\\
$^{18}$Department of Physics and Earth Sciences, University of Ferrara, via 
Saragat 1, 44122 Ferrara, Italy\\
}

\date{Accepted XXX. Received YYY; in original form ZZZ}

\pubyear{2017}

\begin{document}
\label{firstpage}
\pagerange{\pageref{firstpage}--\pageref{lastpage}}
\maketitle

\begin{abstract}
%It should be a single paragraph not more than 250 words (200 words for Letters).
In September 2016, the microquasar Cygnus X-3 underwent a giant radio flare, which was 
monitored for 6 days with the Medicina Radio Astronomical Station and the Sardinia Radio Telescope. Long observations were performed in order to follow the evolution of the flare on a hourly scale, covering six frequency ranges from 1.5 GHz to 25.6 GHz. 
The radio emission reached a maximum of $13.2 \pm 0.7$ Jy at 7.2 GHz and $10 \pm 1$ Jy at 18.6 GHz. Rapid flux variations were observed at high radio frequencies at the peak of the flare, 
together with rapid evolution of the spectral index: $\alpha$ steepened from 0.3 to 0.6 (with $S_{\nu} \propto \nu^{-\alpha}$) within 5 hours. This is the first time that such fast variations are observed, giving support to the evolution from optically thick to optically thin plasmons in expansion moving outward from the core. Based on the Italian network (Noto, Medicina and SRT) and extended to the European antennas (Torun, Yebes, Onsala), VLBI observations were triggered at 22 GHz on five different occasions, four times prior to the giant flare, and once during its decay phase. Flux variations of 2-hour duration were recorded during the first session. They correspond to a mini-flare that occurred close to the core ten days before the onset of the giant flare. From the latest VLBI observation we infer that four days after the flare peak the jet emission was extended over 30 mas.

\end{abstract}


%%%%%%%%%%%%%%%%%%%%%%%%%%%%%%%%%%%%%%%%%%%%%%%%%%

%%%%%%%%%%%%%%%%% BODY OF PAPER %%%%%%%%%%%%%%%%%%

\section{Introduction}

Galactic X-ray binaries with jets are called microquasars, in analogy to the phenomena seen in quasars but on much smaller scales \citep{Mirabel_1999}. Most  microquasars host a stellar-mass black hole as the compact object. They spend most of their time in a dormant state and suddenly enter into periods of outburst activity. Major progress has been made in the understanding of the accretion/ejection connections thanks to multi-wavelength observations. However, the formation of relativistic jets, their composition and exact launching mechanisms are still poorly known. 
%What fraction of the inflow goes into the jets ? 

Discovered by \citet{Giacconi_1967} at the dawn of X-ray astronomy, Cygnus X-3 (Cyg X-3) is
a rare
%the only known 
high-mass X-ray binary consisting of a compact object wind-fed by a Wolf-Rayet star \citep{vanKerkwijk_1996, Fender_1999, Koch-Miramond_2002}. 
%Cyg X-3 is a certain progenitor of a double compact system, after its Wolf–Rayet star explodes as a supernova (Belczynski et al. 2013).
The nature of the compact object is still uncertain, but a black hole seems to be favored considering X-ray and radio emissions \citep[][and references therein]{Hjalmarsdotter_2009, Shrader_2010}. Located in the Galactic plane at a distance of 7--9 kpc \citep{Predehl_2000, Ling_2009, McCollough_2016}, Cyg X-3 has a short orbital period of 4.8 hr \citep{Parsignault_1972}.
It is a reasonably strong persistent radio source with a typical flux of about 100--200 mJy in the quiescent state \citep{Waltman_1996}. 
Flares of various amplitudes are frequently detected, classified as minor or major flares according to the flux density below or above 1 Jy. 
Quenched radio states \citep[< 30 mJy;][]{Waltman_1996} are occasionally observed for Cyg X-3. They are usually followed by major radio flares on a scale of a few days or weeks \citep{Waltman_1994, Waltman_1995}. 
Radio fluxes of 10--20 Jy associated with giant flare events have been measured, with an increasing flux of a factor $\sim 1000$ in a few days \citep{Waltman_1995, Mioduszewski_2001, Miller-Jones_2004, Corbel_2012}. No other X-ray binary has shown such strong and uncommon flux densities up to 20 Jy, which makes Cyg X-3 the brightest X-ray binary at radio frequencies. 
%(McCollough et al. 1999 ???). 
One- and two-sided relativistic jets with a complex structure were clearly resolved during these episodes using the Very Long Array (VLA), the Very Long Baseline Array (VLBA) and the European Very Long Baseline Interferometry Network (e-EVN) \citep{Marti_2001, Mioduszewski_2001, Miller-Jones_2004, Tudose_2007}. 
% two side jets, 0.5-0.8c,  ...
%reaching until 20 Jy at 15 GHz for the brightest ones (). 
%giant radio outbursts with
%strong  evidence  of  jet-like  structures  moving  away  at  relativistic speeds  (Molnar,  Reid  &  Grindlay  1988;  Schalinski  et  al.  1995; Mioduszewski et al. 2001)

X-ray data from Cyg X-3 is more complex compared to other X-ray binaries \citep{Bonnet-Bidaud_1988, Szostek_2008, Koljonen_2010}.
%The X-ray picture of Cyg X-3 is more complex, in particular the behavior of the X-ray spectra which are different from the other X-ray binaries \citep{Bonnet-Bidaud_1988, Szostek_2008, Koljonen_2010}.
%McClintock \& Remillard 2000; Fender et al. 2004
Strong X-ray absorption at low energies is likely associated with the dense wind of the companion star \citep{Szostek_2008}. 
Despite the complex spectral behaviour, 
%it has been
%Deep studies of Cyg X-3 have 
observations have demonstrated that the source exhibits the canonical X-ray states (hard, intermediate, and soft states; \citealt{McClintock_2006}; \citealt{Belloni_2010}) in addition to the very high state and the ultra-soft state \citep{Szostek_2004, Hjalmarsdotter_2009}.
%, and quiescence \citep{Hjalmarsdotter_2009}.
The connections between the X-ray (accretion) and radio (ejections) emission of Cyg X-3 have been widely studied \citep{Watanabe_1994, McCollough_1999, Gallo_2003, Hjalmarsdotter_2008, Szostek_2008b, Zdziarski_2016}. In particular, giant radio flares correspond to the transition from the ultra-soft state to a harder X-ray state \citep{Koljonen_2010}. 
%The relationship between X-ray and radio emission is deeply studied .

Cyg X-3 was also the first microquasar detected in gamma rays with AGILE and Fermi/LAT \citep{Tavani_2009, Fermi_2009}, providing unique insight into the particle acceleration up to GeV energies during the ejection.
Gamma-ray emission was detected by Fermi-LAT during the 2011 giant flare episode \citep{Corbel_2012}. 
The high-energy emission (> 100 MeV) corresponds to transitions in and out of ultra-soft X-ray state.
%The high-energy emission (> 100 MeV) corresponds to the transitions (beginning and end) of the ultra-soft X-ray state (radio quenched).
%The multi-wavelength observations testified that 
%Cyg X-3 represents an exceptional target that offers the possibility to better understand the relationship between accretion state and jet launching, in particular in the extreme cases of giant outbursts.  
%the multi-wavelength observations testify the strong connection between very-low accretion rates and strong relativistic jet production.

The Sardinia Radio Telescope (SRT\footnote{\url{www.srt.inaf.it}}) has carried out a large-monitoring program of several X-ray binaries during the Early Science Program (ESP\footnote{\url{www.srt.inaf.it/astronomers/early-science-program-FEB-2016/}}), from February to July 2016 (PI: Egron). The weekly monitoring performed at 7.2 and 22 GHz has shown that Cyg X-3 was in the quiescent state during this period. 
%The SRT frequencies are complementary to the RATAN-600 that performed a daily monitoring of the source. 
A quenched radio state was detected by the RATAN-600 on 23--25 August 2016 \citep{Trushkin_2016ATel9416}, five and a half years after its last quenched episode. The ultra-soft X-ray state was confirmed by Swift/BAT\footnote{see \citealt{Krimm_2013} for a description of the Swift/BAT transient monitor},
%\footnote{http://swift.gsfc.nasa.gov/results/transients/CygX-3/}, 
which registered a strong decrease of a factor 25 of the hard X-ray emission in the 15--50 keV band in less than 10 days, from 15 August to 24 August. The AGILE-GRID detector revealed gamma-ray emission above 100 MeV consistent with the position of Cyg X-3 on 28--30 August \citep{Piano_2016}. 
%The gamma-ray emission was then confirmed by 
%Fermi/LAT, cf notre ATel du 15/16 septembre Cheung et al.  ATel 9502
The strong increase of the radio flux occurred from 14 September, three weeks after the transition to the quenched radio state \citep{Trushkin_2016ATel9416, Trushkin_2016ATel9501}. A gamma-ray flare was detected by Fermi/LAT the same days, on 15--16 September \citep{Cheung_2016}.
%Fermi/LAT, cf notre ATel du 15/16 septembre Cheung et al.  ATel 9502

In this paper, we present single-dish and Very Long Baseline Interferometry (VLBI) observations of Cyg X-3 corresponding to the 2016 September giant flare episode. 
We present the details of the observations and data reduction in Section\,2, the evolution of the flux density and spectral index in Section\,3, and discuss the results we obtained, in particular 
the jet characteristics and morphologies associated with the mini and giant flares in Section\,4.
%(ultra-soft X-ray state). 

%2016-08-23 12:00:00 => 57623 cf light curve swift

%Evolution of Cyg X-3 and catch the source in the expected giant flare.

%Single-dish observations were performed with SRT and Medicina (Atels) : monitoring the flux density during the giant flare, at different frequencies, for several hours per day
%ToO VLBI program triggered (SRT, Medicina, Noto, Torun, Yebes) at different period of the flare.
%Gamma-rays : AGILE observations (Nature  2009; Corbel 2012 ; Atel 2016) ?

\section{Observations and data reduction}

We initially planned to perform VLBI observations to catch the source during the rising and declining phases of the flare in order to track the evolution of the relativistic jets.
Due to the difficulty to trigger VLBI observations in a very short time (considering the availability of the antennas), we conducted single-dish observations at the moment of the peak of the flare \citep{Egron_2016a}.
The recently commisioned 64\,m-SRT \citep{Bolli_2015, Prandoni_2017} participated in both single-dish and VLBI observations, together with the 32\,m-Medicina radio telescope. 
VLBI observations included SRT, Medicina, Noto, Torun, Yebes and Onsala in order to provide a larger coverage of the $(u,v)$-plane.

\subsection{Single-dish observations}

A Target-Of-Opportunity program for Cyg X-3 was triggered by the Italian Medicina Radio Astronomical Station and SRT single dishes. The outburst was followed
% in order to follow the evolution of the giant burst, 
from 17 to 23 September 2016. Both antennas were equipped with the same control software
%Enhanced Single-dish Control System (ESCS; DISCOS project) 
designed to optimize single-dish observations \citep{Orlati_2016}. 
The frequency agility (switching the observing frequencies in only a few minutes) coupled with
the different SRT and Medicina receivers  
%The receivers of SRT and Medicina being complementarity  cou
enabled a large frequency coverage of Cyg X-3 during the flare, from 1.5 GHz to 25.6 GHz.

%SRT observations were carried out at 1.55 GHz, 7.24 GHz and 22.66 GHz 
SRT observations were carried out at 1.5 GHz (L-band), 7.2 GHz (C-band), and 22.7 GHz (K-band) 
using the Total Power and SARDARA (Melis et al. in prep.) backends in piggy-back mode. Observations consist of rectangular and perpendicular On-The-Fly (OTF) maps performed in the Right Ascension (RA) and Declination (DEC) directions (forming a Greek cross map), at constant velocity ($4\arcmin$/sec). The dimensions of the maps were chosen according to the beam size (see Table~\ref{tab:beam-size}) at the observed frequencies: $1.5^\circ \times 0.6^\circ$ at 1.55 GHz, $0.5^\circ \times 0.12^\circ$ at 7.2 GHz and $0.2^\circ \times 0.05^\circ$ at 22.7 GHz. This method has been applied during the Early Science Program dedicated to the monitoring of X-ray binary systems. The observing strategy has the advantage of providing a direct image of the sources in the vicinity of the target and a better estimate of the flux density. The data analysis was accomplished with the SRT Single-Dish-Imager (SDI; \citealt{Egron_2016b}), a software designed to perform automated baseline subtraction, radio interference rejection, and calibration. The spectral flux density of the target was reconstructed by observing three calibrators (3C286, 3C295 and 3C48) at all frequencies, by applying the values and polynomial expressions proposed by \citet{Perley_2013}.

Cyg X-3 was observed with Medicina at 8.5 GHz (X-band), 18.6 GHz and 25.6 GHz (K-band), using the OTF cross-scan technique in RA and DEC directions. In X-band, the bandwidth was 680 MHz with a scan length and velocity of $0.6^\circ$ and $2.4\arcmin$/sec respectively, while in K-band we selected a bandwidth of 1200 MHz, scans of $0.2^\circ$ length and a scan velocity of $0.8\arcmin$/sec.
%\textit{(size?, additional information Simona)}. 
%******
%X-band 
%******
%Lunghezza subscan: 0.6°
%Velocità di scansione: 2.4 °/min
%Larghezza di banda: 680 MHz
%Campionamento: 40 ms
%
%******
%K-band 
%******
%Lunghezza subscan: 0.2°
%Velocità di scansione: 0.8 °/min
%Larghezza di banda: 1200 MHz
%Campionamento: 40 ms
We applied gain curve and pointing offset corrections to the measurements. Additional opacity and atmospheric corrections were added in the case of K-band data (18.6 GHz and 25.6 GHz). The flux calibration was performed with observations of 3C286, 3C48 and NGC7027. The flux of 3C286 was calculated according to \citet{Perley_2013}, while the fluxes of the other calibrators were reckoned on the basis of \citet{Ott_1994}.
The single-dish observations are reported in Table~\ref{tab:single-dish-summary}.

\begin{table}
	\centering
	\caption{Beam sizes at the observed frequencies using  SRT and Medicina as single-dish radio telescopes.}
	\label{tab:beam-size}
	\begin{tabular}{ccc} % four columns, alignment for each
		\hline
Radio		&  Frequency& Beam size \\
Telescope	&  (GHz)	& (arcmin) \\
		\hline
SRT		& 1.5		& 12.2 \\  %12.5
		& 7.2 	& 2.63 \\ %2.66
		& 22.7 	& 0.816 \\ %0.895
		\hline
Medicina 	& 8.5		& 4.55	\\
		& 18.6 	& 2.08 \\
		& 25.6	& 1.51 \\
		\hline
	\end{tabular}
\end{table}

%Table 1: Single-dish observations of Cyg X-3.
\begin{table}
	\centering
	\caption{Single-dish observations of Cyg X-3 performed with Medicina at 8.5, 18.6 and 25.6 GHz, and with SRT at 1.5, 7.2 and 22.7 GHz.}
	\label{tab:single-dish-summary}
	\begin{tabular}{cccc} % four columns, alignment for each
		\hline
		Obs date 	 &	 MJD	 & Frequency & Flux density\\
			     	 &                & (GHz) 	 &  (Jy) \\
		\hline
		17 Sept 2016 & 57648.86 & 8.5	 & 13.1 $\pm$ 0.4 \\
     				    & 57648.88 &    	 & 12.9 $\pm$ 0.4 \\
				    & 57648.90 &   	 & 13.0 $\pm$ 0.4 \\
				    & 57648.92 &   	 & 12.9 $\pm$ 0.4 \\
				    & 57648.95 & 18.6   & 7.5 $\pm$ 0.8 \\  
				    & 57648.97 & 25.6   & 7.0$\pm$ 0.7 \\  
%		\hline
		19 Sept 2016 & 57650.62 & 7.2	 & 12.0 $\pm$ 0.6 \\
				    & 57650.65 & 8.5	 & 12.9 $\pm$ 0.4 \\
				    & 57650.67 & 		 & 12.9 $\pm$ 0.4 \\
				    & 57650.67 & 7.2 	 & 12.6 $\pm$ 0.6 \\
				    & 57650.70 &		 & 13.1 $\pm$ 0.7 \\
				    & 57650.70 &	18.6	 & 8.9 $\pm$ 0.9 \\
				    & 57650.71 &	7.2	 & 12.8 $\pm$ 0.6 \\
%				    & 57650.72 & 22.7  & 7.9 $\pm$ 0.8  \\
				    & 57650.72 & 22.7  & 9.5 $\pm$ 0.9  \\
				    & 57650.73 & 18.6    & 10 $\pm$ 1  \\
				    & 57650.77 & 7.2    & 13.2 $\pm$ 0.7 \\
				    & 57650.88 &		  & 12.8 $\pm$ 0.6 \\
				    & 57650.91 &		  & 12.9 $\pm$ 0.6 \\
				    & 57650.91 & 25.6    & 6.1 $\pm$ 0.6 \\
				    & 57650.93 &            & 5.6 $\pm$ 0.6 \\
%				    & 57650.95 & 22.7  & 5.3 $\pm$ 0.5 \\
				    & 57650.95 & 22.7  & 6.4 $\pm$ 0.6 \\
%		\hline
		20 Sept 2016 & 57651.54 & 7.2	 & 11.4 $\pm$ 0.6 \\
				    & 57651.61 & 8.5     & 12.1 $\pm$ 0.4 \\
				    & 57651.63 & 7.2     & 11.8 $\pm$ 0.6 \\
				    & 57651.65 & 18.6    & 7.2 $\pm$ 0.7 \\
%				    & 57651.66 & 22.7    &  (3.59) ? \\
				    & 57651.68 & 25.6    & 7.3 $\pm$ 0.7 \\
%		\hline
		21 Sept 2016 & 57652.67 & 7.2	 & 6.9 $\pm$ 0.3 \\
%				    & 57652.72 & 22.7    & 3.0 $\pm$ 0.3 \\  
				    & 57652.72 & 22.7    & 3.6 $\pm$ 0.4 \\     
				    & 57652.76 & 1.5      & 14.4 $\pm$ 0.5 \\  % central pix map Jy/beam 
%    				    & 57652.76 & 1.5      & 15.9 $\pm$ 0.5 \\ 2 beams
%		\hline
		22 Sept 2016 & 57653.97 & 8.5      & 3.2 $\pm$  0.1\\
				    & 57653.99 &	 	 & 3.1 $\pm$  0.1 \\
%                \hline
		23 Sept 2016 & 57654.00 & 8.5      & 3.1 $\pm$  0.1\\
		\hline
	\end{tabular}
\end{table}

\subsection{VLBI observations}

\begin{figure*}
   \centering
   \includegraphics[width=0.72\textwidth]{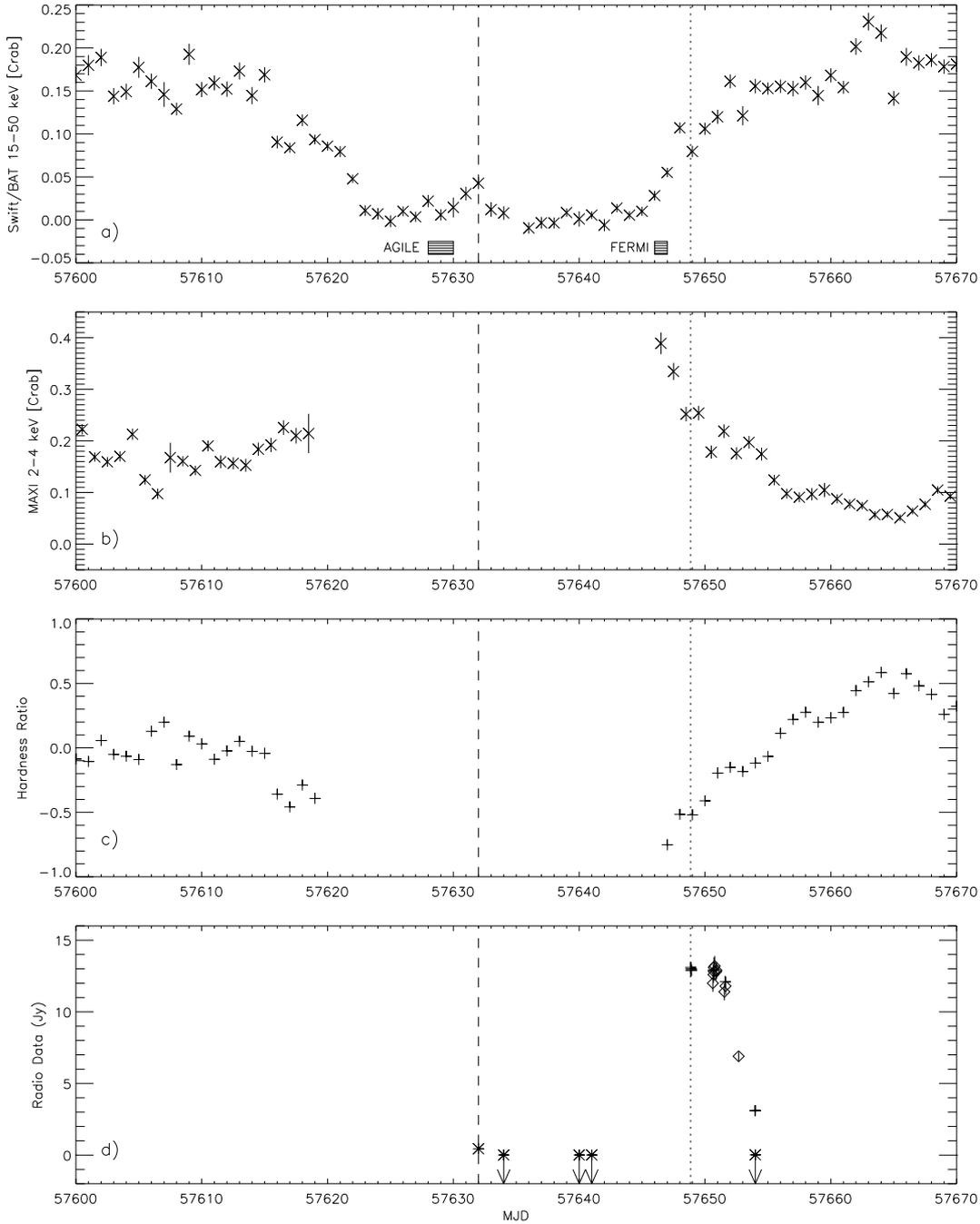}
    \caption{
From the top to the bottom: a) Swift/BAT 15--50 keV (Hard X-rays) light curve of Cyg X-3 in Crab units from 31 July to 09 October 2016. The gamma-ray detections with AGILE and Fermi/LAT are also reported; b) MAXI 2--4 keV (Soft X-rays) light curve in Crab units; c) Hardness ratio corresponding to (Hard-Soft)/(Hard+Soft); d) Single-dish (cross: 8.5 GHz; diamond: 7.2 GHz) and VLBI data at 22 GHz (asterisk). The dashed line and the dotted line indicate the first VLBI observation and the first single-dish observation, respectively. The vertical arrows represent the VLBI upper limits at $5\sigma$ confidence.}
    \label{fig:lc-tt}
\end{figure*}

%In the context of the approved IRA proposal $18-16$ ''VLBI ToO observations of neutron star and black hole X-ray binaries'', 
We triggered five VLBI observations on Cyg X-3, from the quenched radio state until the end of the strong flare episode. Our aim was to catch the target in the rising and fading phases of the predicted giant flare \citep{Trushkin_2016ATel9416} in order to follow the evolution of the relativistic jets.
The first two VLBI observations were triggered on 1 and 3 September 2016,
%(from 16 UT until 05 UT the day after), 
right after the first increase of the hard X-ray emission recorded by the Swift/BAT on 29 August 2016. It turned out that the X-ray flux decreased three days later and the giant flare did not occur at that time. 
We triggered other VLBI observations a few days later, on 9 and 10 September 2016, after a second increase of the X-ray flux on 8 September. 
The last observation was carried out on 23 September, during the declining phase of the giant radio flaring episode. The observing sessions are indicated in the bottom panel of the Figure~\ref{fig:lc-tt}.
% in addition to the soft and hard X-ray light curves obtained with MAXI and Swift/BAT.
%The VLBI observations are indicated in Fig... .
%to the Swift/BAT light curve with a dashed line.

The VLBI observations were performed at 22 GHz with the following radio telescopes, according to their availability: SRT (Sr), Medicina (Mc), Noto (Nt, 32\,m, Italy), Torun (Tr, 32\,m, Poland), Yebes (Ys, 40\,m, Spain) and Onsala (On, 20\,m, Sweden). 
We alternately observed Cyg X-3 and the calibrators 3C345, BL Lac, J2007+4029, and J2015+3710, resulting in sessions of 7--15.5 hrs each. 
%The data were processed with the Medicina correlator (ref).
The data were processed with the DiFX correlator \citep{Deller_2011} installed and operated in Bologna.
The VLBI analysis was performed with the Astronomical Image Processing System \citep[AIPS;][]{Greisen_2003}. 
The calibrator J2007+4029 was used to perform the phase referencing. 
The phase-referencing cycle was 4\,min: 2.5\,min on the target and 1.5\,min on phase-cal.
Since the calibrator-target separation was quite large ($d=4.7^\circ$), and the
region is subject to significant scatter, we also observed the check source
J2015+3710 every 30\,min. This source was also phase-referenced to
J2007+4029 and it was used to confirm that the phase solutions were
transferred correctly.
%Fringe fitting was performed separately for Cyg X-3 and the calibrator sources.
We also tried to fringe-fit the Cyg X-3 data directly, in order to estimate
its signal-to-noise ratio in each 2.5\,min long scan. However, these results
were rather used for data quality assessment than for further analysis.

\section{Results}

\subsection{Single-dish monitoring of the giant flare}

Medicina and SRT provided the monitoring of Cyg X-3
from 17 September to 23 September 
%during the peak of the flare. 
%at 1.55, 7.24, 8.52, 18.6, 22.66 and 25.6 GHz, 
at 1.5, 7.2, 8.5, 18.6, 22.7 and 25.6 GHz, which was complementary to the daily monitoring of the source performed with the RATAN-600 \citep[covering the 2.3, 4.6, 8.2, 11.2, 21.7 GHz frequencies;][]{Trushkin_2016}. 
%The detail of the single-dish observations of Cyg X-3 is reported in Table~\ref{tab:single-dish-summary}. 
Our observations are summarized in Table~\ref{tab:single-dish-summary} and represented in Fig.~\ref{fig:lc-tt}, together with the X-ray light curves obtained with data daily averages from Swift/BAT at 15--50 keV and MAXI at 2--4 keV in Crab units\footnote{1 Crab = 0.22 ct/cm$^{2}$/sec for the Swift/BAT rate and 1 Crab = 1.67 ct/cm$^{2}$/sec for the MAXI rate.}. 
The hardness ratio obtained from Swift/BAT and MAXI data is also reported despite the lack of MAXI data during most of the ultra-soft X-ray state, even though present in the Swift/BAT light curve.
%Moreover, the long-sessions observations provided with the Italian radio telescopes allow us to infer variation of the radio emission on hour scale, as the case of the 19 September.

\begin{figure}
   \centering
   \includegraphics[width=0.95\columnwidth]{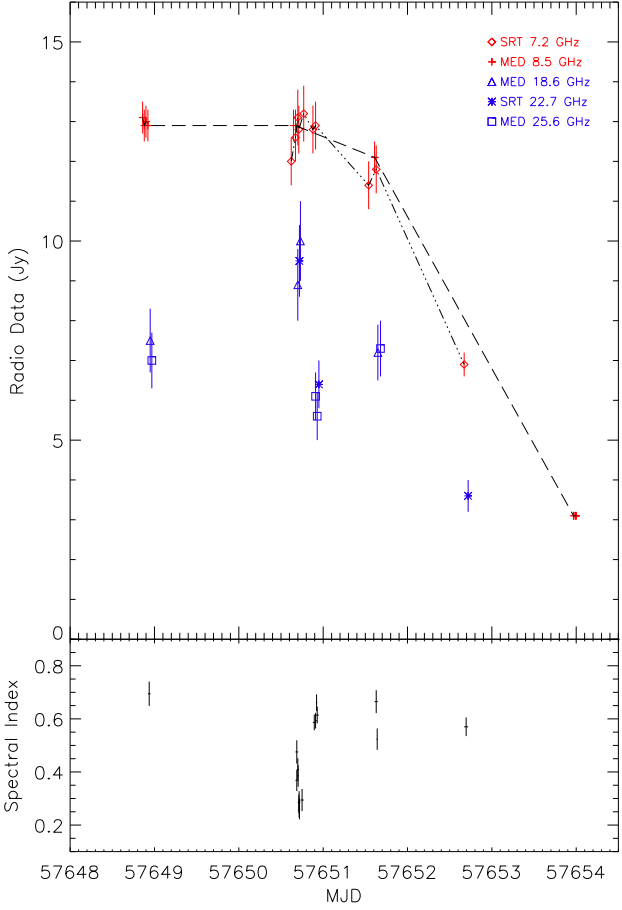}
   \vspace*{6mm}
    \caption{Top panel: Medicina and SRT observations covering the peak of the giant flare (see Table~\ref{tab:single-dish-summary} for more details). Lower frenquencies are indicated in red while higher frequencies are in blue.
%Cross and diamond symbols indicate 8.5 and 7.2 GHz data, respectively. Triangles, squares and asterisks correspond to 18.6, 25.6 and 22.7 GHz observations, respectively. 
Bottom panel: Evolution of the spectral index $\alpha$.}
    \label{fig:single-dish}
\end{figure}

\begin{figure}
   \centering
   \includegraphics[width=0.85\columnwidth]{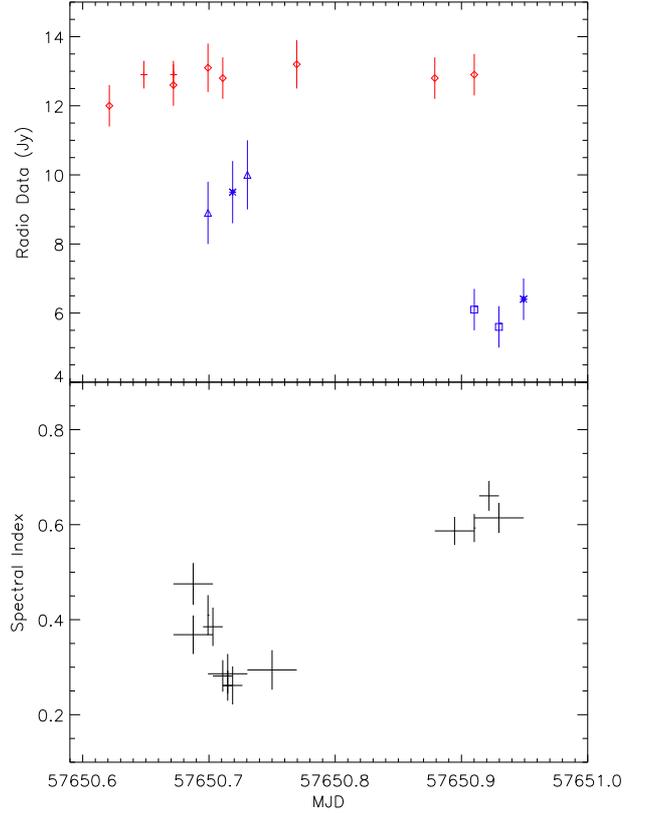}
   \vspace*{6mm}
    \caption{Zoom of the light curve and spectral index evolution presented in Fig.~\ref{fig:single-dish} during the maximum of the giant flare peak, on 19 September 2016. 
%Cross and diamond symbols indicate 8.5 and 7.2 GHz data, respectively. Triangles, squares and asterisks correspond to 18.6, 25.6 and 22.7 GHz observations, respectively.
}
    \label{fig:zoom-spec-ind}
\end{figure}

\begin{figure}
    \centering
    \includegraphics[width=\columnwidth]{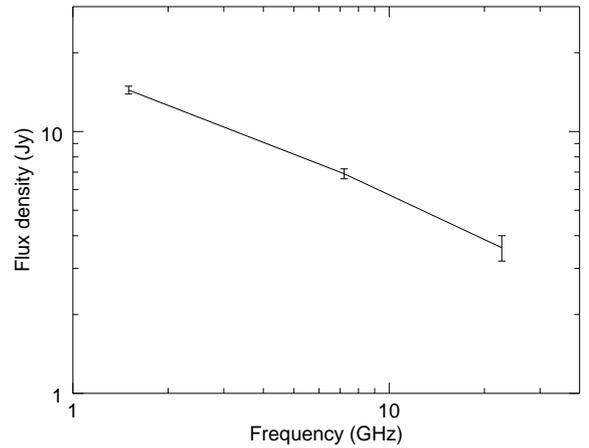}
    \caption{Optically thin spectrum obtained from SRT data at 1.5, 7.2 and 22.7 GHz two days after the maximum of the giant flare.}
    \label{fig:21sept-spectrum}
\end{figure}

\begin{figure*}
   \includegraphics[width=1.0\textwidth]{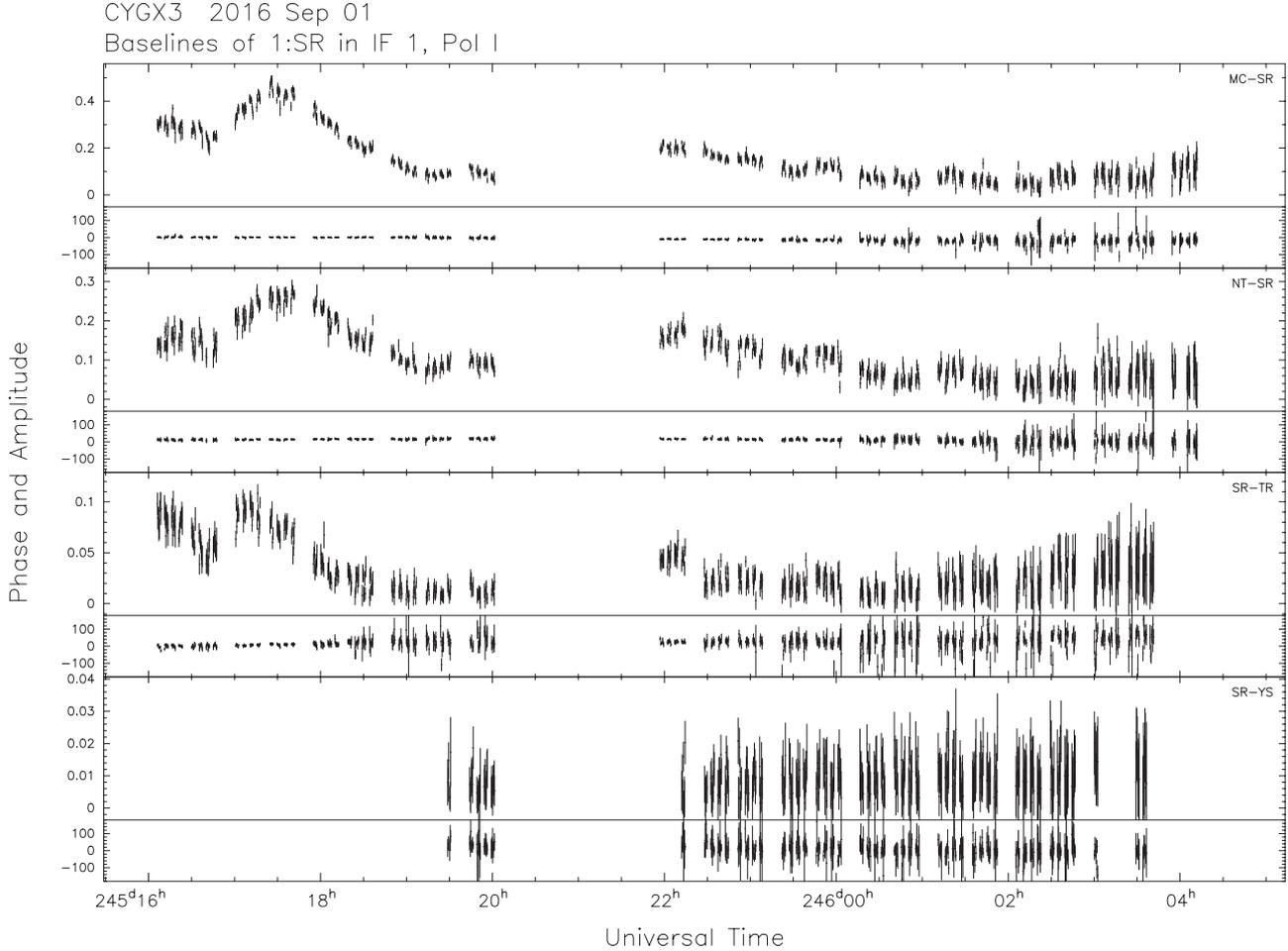}
    \caption{Cyg X-3 visibility amplitude and phase vs. time for baselines to
SRT for the 1 September 2016 (MJD 57632) observations at 22 GHz. 
Four pairs of panels are shown; from top to bottom: Sr-Mc, Sr-Nt,
Sr-Tr, Sr-Ys; in each pair, the top panel shows amplitudes (in Jy) and the
bottom phases (in degrees). Data are missing at around UT 21 for these
baselines, due to SRT elevation limits (source transiting at zenith), but
are in general present on other baselines. Note that the first part of the observation was missed by Yebes due to technical problems.} 
%Light curve of Cyg X-3 obtained at 22 GHz during the VLBI observation carried out on 1 September 2016. A variability is observed within two hours.}
    \label{fig:vplot-all}
\end{figure*}

\begin{figure}
   \includegraphics[width=\columnwidth]{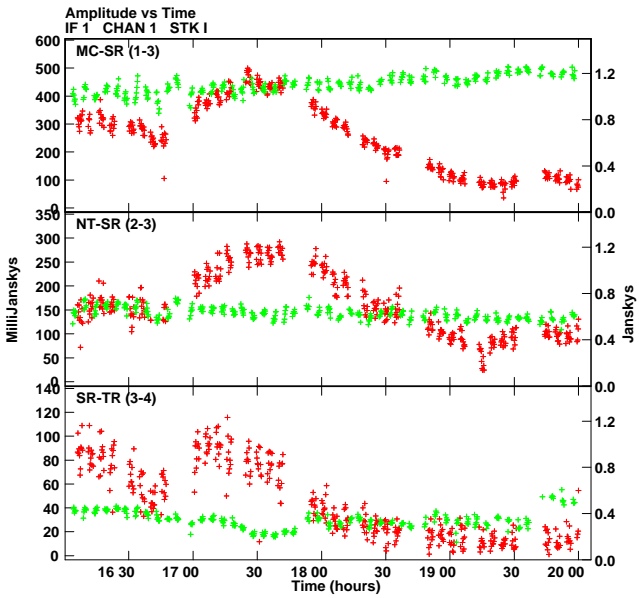}
    \caption{
Visibility amplitudes of Cyg X-3 (red) and J2007+4029 (green)
%and J2015+3710 (blue) 
during the four first hours of the 1 September 2016
observation (MJD 57632). Variability is observed on the Mc-Sr, Nt-Sr, Sr-Tr baselines  for 
Cyg X-3 within two hours whereas the comparison source remains stable.}
%The scale on the y-axis is between 0 and 4 Jy for J2015+3710 and between 0 and 1.5 Jy for J2007+4029, on all baselines.
    \label{fig:VLBI-1sept}
\end{figure}

The giant radio flare reached its highest brightness on 19 September 2016 (MJD 57650.7). While the flux density remained quite stable ($\sim 13$ Jy) at 7.2--8.5 GHz during the peak maximum, 
%The maxima of the flux density $\sim 13$ Jy at 7.2 GHz and $\sim 10$ Jy at 18.6 GHz were reached on 19 September 2016 (MJD 57650.7), which confirmed the designation of ''giant flare''. 
%The flux density was quite stable during these days, with the indication of some variations on the scale of a few hours (see Fig~\ref{fig:single-dish}). 
%While the flux at $7.2-8.5$ GHz remained quite stable during the peak maximum, 
we observed flux variations on the timescale of a few hours in K-band, as shown in  Fig.~\ref{fig:single-dish} and \ref{fig:zoom-spec-ind} (note that all error bars are at the $1\,\sigma$ level). The 18.6, 22.7 and 25.6 GHz observations performed independently (observational and data reduction techniques) with Medicina and SRT are perfectly consistent, demonstrating a decrease of the flux within less than 5 hours before it increased again the day after, on 20 September.
On the other hand, the radio emission at 7.2--8.5 GHz severely weakened 
%from $20-21$ September for most of the frequencies.
from 20 September; a reduction of $\sim 9$ Jy is detected within 2.4 days.
Observations carried out with the SRT on 21 September at 1.5, 7.2 and 22.7 GHz indicated that the
ejection is optically thin,
%optical depth was negligible, 
as shown in Figure~\ref{fig:21sept-spectrum}.

The evolution of the spectral index $\alpha$ (with $S_{\nu} \propto \nu^{-\alpha}$) is shown in Figure~\ref{fig:single-dish}. 
%The spectral index was calculated on time intervals of 1 hr, considering two groups of frequencies, 7.2--8.5 GHz and 18.6--25.6 GHz.
The spectral indexes were calculated by considering all couples of flux density measurements available at low and high frequencies (7.2--8.5 GHz and 18.6--25.6 GHz) within 1 hr from each other.
Spectral index errors were derived from error propagation of flux density errors for each 
couple. The time tag of the reported spectral index values corresponds to the mid time between the epochs of each flux density measurement couple. 
The error bars on the x-axis (time) reflects the epoch 
separation for each couple of flux density measurements.
%The power-law spectral indeces reported in fig. 2 and 3 are obtained from  all couples of flux density measurements available at different  frequencies and obtained within 1 hour from each other. Spectral index  errors are derived from error propagation of flux density errors for each  couple. The time tag of reported spectral indeces are obtained by  considering the mid time between the epochs of each flux density  measurement couple. The error bars on the x-axis (time) reflects the epoch separation for each couple of flux density measurements.
We clearly observed a spectral steepening  from  $\alpha =$
%$0.4 \pm 0.1$ to $0.62 \pm 0.05$ within $\sim5$ hrs
$0.34 \pm 0.08$ to $0.61 \pm 0.03$ within $\sim5$ hrs
%$0.70 \pm 0.05$ to $0.30 \pm 0.05$ 
at the moment of the peak maximum of the flare.
%It decreased from 0.7!!! to 0.5!!!, then increased to come back to the previous!!! value. 
%We considered time intervals of 5 hours to calculate $\alpha$ considering 
In Figure~\ref{fig:zoom-spec-ind}, we show a zoom of the light curve and spectral index  evolution for the 
%at the moment of the maximum of the flare, 
observation of 19 September 2016 (MJD 57650).

\subsection{VLBI results}

Compact (milliarcsecond scale) radio emission was detected from Cyg X-3
during the first VLBI observation, on 1 September 2016. This detection was confirmed
both for the phase-referenced visibilities, whose amplitude and phase
showed well-defined coherence, and from a run of fringe fitting directly to
the source itself, which produced good solutions for all the intervals with
valid data.

%Table 2: VLBI observations of Cyg X-3
\begin{table*}
	\centering
	\caption{VLBI observations of Cyg X-3 performed at 22 GHz with SRT, Medicina, Noto, Torun, Yebes and  Onsala. MJD start and end indicate the beginning and the end of each session while the effective time is related to the observation time on Cyg X-3. Upper limits are given at $5 \sigma$ when the source was not detected at the known position.}
	\label{tab:vlbi-summary}
	\begin{tabular}{cccccccc} % four columns, alignment for each
		\hline
Radio telescopes  & Obs. date &	 MJD	 &  MJD & Flux density \\ % &    Beam  \\%	& rms\\
			 &                 & 	(start) &  (end)&  (mJy)	\\ %&     (mas)	 \\%& (mJy/beam)\\
%Eff. time  (hrs) & 
		\hline
Sr, Mc, Nt, Tr, Ys &1 Sept 2016& 57632.67 &  	57632.79$^{\star}$ &  440	\\%& 		\\%&   \\
			&		    &  57632.79$^{\star}$ &  57633.21  & 250 	\\%&		&  \\
Sr, Mc, Nt, Tr, Ys & 3 Sept 2016 & 57634.67 & 57635.21  & $< 8$	\\%&		\\%& \\ 
Sr, Mc, Nt, Tr      & 9 Sept 2016 & 57640.67  & 57641.19	& $< 8$	\\%&		\\%& \\	 	
Sr, Mc, Nt, Tr	& 10 Sept 2016 & 57641.54 &  57641.83 & $< 8$	\\%& 		\\%& \\	
Mc, Nt, Tr, On & 23 Sept 2016 & 57654.50 & 57655.15  & $< 20$	\\%& 		\\%& \\		
		\hline
	\end{tabular}
              \\
  	    \small
   	   $^{\star}$ The observations are split to report the different flux density values but they correspond to the same run.
%	\begin{tablenotes}
  %	    \small
 %  	   $^{\star}$ Upper limits at $3 \sigma$ when Cyg X-3 was not detected at the known position.
   %	 \end{tablenotes}
\end{table*}

In Fig.~\ref{fig:vplot-all}, we show visibility amplitude and phase versus time
for all the baselines to SRT. Other baselines show similar behaviour but
with increased scatter due to lower baseline sensitivity. We note a
variable total flux density: the amplitudes are in general higher in the
first half of the session than in the second. Moreover, there is clear
variability on hour-scale in the initial four hours of the observation. 
We exclude the possibility that this variability is instrumental: we present in Fig.~\ref{fig:VLBI-1sept} the visibility amplitudes (Mc-Sr, Nt-Sr, Sr-Tr baselines) for Cyg X-3 and J2007+4920 
%and J2015+3710 
for the first four hours of observations. 
The comparison source shows the stability of the
system, which allows us to infer that the amplitude variation is intrinsic
to Cyg X-3. Moreover, the variations follow the same pattern on
baselines of very different orientations in the $(u,v)$-plane (including
those not shown: Mc-Nt, Mc-Tr, Nt-Tr). This behaviour indicates that the amplitude
variability is not due to the source structure but rather to intrinsic flux
density variations.

\begin{figure*}
    \centering
    \includegraphics[width=0.8\textwidth]{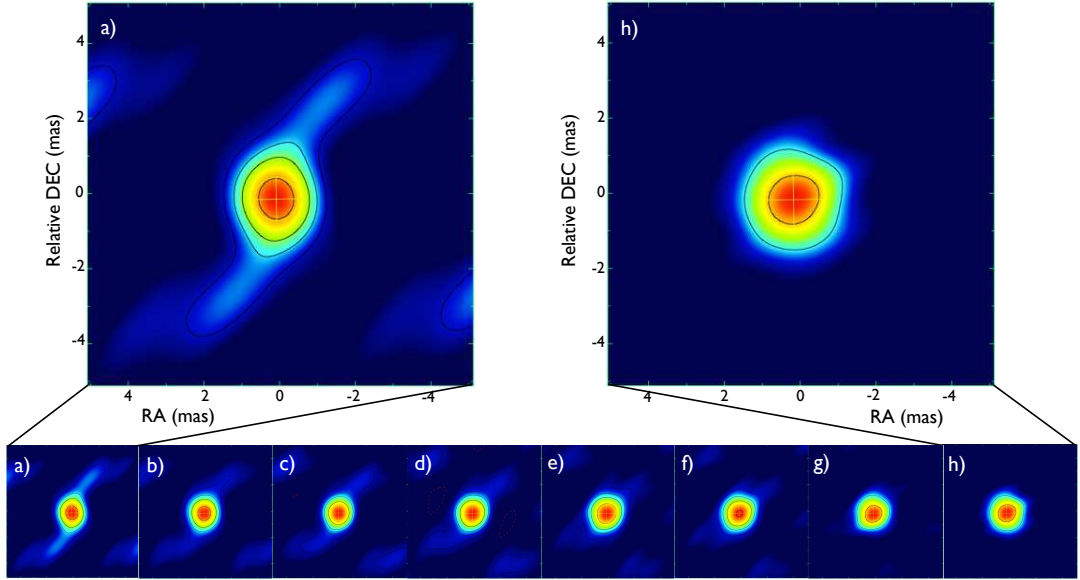}
    \caption{Evolution of the size of the emitting component during the first 4 hours of the VLBI observation performed on 1 September 2016 (MJD 57632). The images were obtained on short time intervals of 15 min by convolving the best-fit circular Gaussian with a circular beam of 1.5 mas HPBW.
A sequence of images from a) to h) is represented, with an interval of 30 min between two consecutive images for clarity.
%and are represented on the sequence from a) to h) each 30 min for clarity. 
The first (a) and last (h) images correspond to 16:15 UT and 20:00 UT, respectively. 
%the contours are the only information about the intensity
}
    \label{fig:jet-1sept-size}
\end{figure*}

\begin{figure}
    \centering
    \includegraphics[width=1.0\columnwidth]{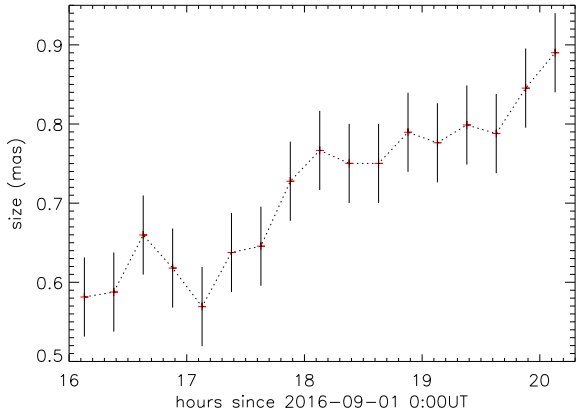}
    \caption{Radius in mas of the emitting component versus time corresponding to Fig.\,\ref{fig:jet-1sept-size}. The size increased within 4 hours, which indicates a signature of the expansion of the emitting region.}
    \label{fig:plot-1sept-size}
\end{figure}

The short-scale time variability 
%denies us the possibility to obtain
prevents us from obtaining a meaningful average image for the whole observation, as the imaging algorithms in interferometry produce a Fourier transform of the
visibility amplitude and phase in the $(u,v)$-plane to the brightness in the sky
combining all visibilities together. Different portions of the $(u,v)$-plane are therefore
sampled at different times. When the source varies during the duration
of the observation, the transform of the $(u,v)$-plane consequently becomes ill-posed. For
this reason, it is not possible to produce a meaningful image of the source
from the entire dataset during our VLBI observation.
A possibility would be to produce images for short time intervals, during
which we can assume that variability plays a negligible role. However, in
this case we can only use the visibilities acquired during that time
interval, which means that we only sample a small portion of the $(u,v)$-plane.
As a result, the image quality becomes very poor (in particular for our
sparse array of only 4--5 stations).
This is the reason why we resorted to model-fitting in the $(u,v)$-plane
rather than to images. We model-fit the visibility within each
time bin of 15\,min by using a circular Gaussian 
%by convolving the best-fit circular Gaussian with a circular beam of 3.0 mas HPBW
model of 1.5 mas HPBW (the beam would naturally be elliptical; note that the real beam size during this observation was 1.50 mas $\times$ 0.97 mas) in which we let position,
amplitude, and width free to vary.  
It is therefore possible to study the evolution
of the size of the emitting component, at least during the first 4 hours of the observation
when we had a better signal-to-noise ratio to constrain the fit. 
The size of the emitting region increased with time from 0.6 to 0.9\,mas radius, as shown in Fig.\,\ref{fig:jet-1sept-size} and \ref{fig:plot-1sept-size}. 
%The expansion from $\sim 1.2$ mas to $\sim 1.8$ mas 
%a signature of the expansion of the emitting region.
% convert to linear sizes and velocities
After 20.00\,UT on 1 September 2016, the size becomes smaller. This shrinking is most likely an artefact
due to the low signal-to-noise-ratio. 
Another intriguing possibility is that the beam position angle is rotating from a direction aligned with the jet axis to a transverse one, so we see a narrower size. 
%But I do not think there is any reliable way to test this.

\begin{figure}
    \centering
    \includegraphics[width=1.0\columnwidth]{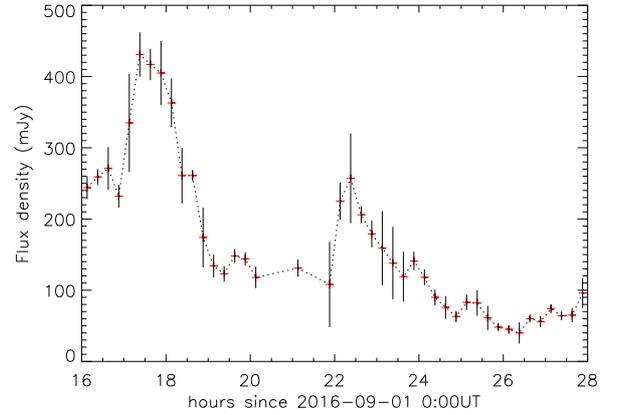}
    \caption{VLBI light curve obtained on 1 September 2016 (MJD 57632) showing a short-scale time variability, with the evidence of two peaks of different flux density during a mini-flare. The general trend of the off-peak measurements indicates a decrease of the flux during the observation.}
    \label{fig:lc-vlbi-15}
\end{figure}

%In parallel, 
%given the good signal-to-noise ratio for the entire duration of the session, we have
%been able to divide the data in short (15\,min) time bins and analyse
%each one of them. W
We have created both clean images and visibility data
models for each time bin of 15\,min and determined the target's flux density using both
methods. The resulting overall trends are entirely consistent.
% Since the image quality is rather limited with short time bins and only few stations,
%we consider in our analysis and 
We report in Figure~\ref{fig:lc-vlbi-15} the numbers provided by
visibility model fits. 
Two small peaks of $\sim 440$ mJy and $\sim 250$ mJy flux density are clearly visible in the light curve, both with a 2-hour duration.
We catched a mini-flare of Cyg X-3 instead of the expected giant flare event.
%Note that this was not the giant flare but rather a pre-flare, so the amplitude scale is actually much smaller than what we expect in the present event. 
The flux density variability on sub-hour scales is however clearly present and it is intrinsic to the source, as demonstrated in Figure~\ref{fig:VLBI-1sept}.

%\begin{figure}
%    \centering
%    \includegraphics[width=\columnwidth]{21sept2016.eps}
%    \caption{Optically thin spectrum obtained from SRT data at 1.5, 7.2 and 22.7 GHz two days after the maximum of the giant flare.}
%    \label{fig:21sept-spectrum}
%\end{figure}

%The following VLBI observations triggered on 3, 9 and 10 September 2016 did not provide any detection of Cyg X-3. Upper limits are reported in Table~\ref{tab:vlbi-summary}. 
The following VLBI observations triggered on 3, 9, and 10 September 2016
did not provide any detection of Cyg X-3 (upper limit of 8 mJy at $5\sigma$), as confirmed by the analysis in both the visibility and the image domains. In the visibility
domain, after transferring phase solutions from the calibrator, we tried to
fringe fit the source visibilities, finding no solution. 
After directly imaging the phase-referenced data,  we did not find any significant peak in the image plane. Upper limits are reported in Table~\ref{tab:vlbi-summary}. Conservatively,
we determined upper limits from the lack of fringe detections at $5\sigma$
significance in 2.5 minutes on the most sensitive baseline, based on the
EVN online calculator\footnote{{\url{
http://www.evlbi.org/cgi-bin/EVNcalc.pl}}}. This is a more conservative
approach than reporting the image rms noise. The latter is typically much
lower but could underestimate the true source flux density in case
of (1) variability on time scales shorter than the observation duration or
(2) signifcant coherence losses due to imperfect phase referencing.
These results are in agreement with the light curve obtained with the RATAN-600 at lower frequencies; the flux densities are between 10 and 40 mJy at 2.3, 4.6, 8.2, and 11.2 GHz \citep{Trushkin_2016}, which indicates a quenched radio state. The hard X-ray emission (Swift/BAT light curve) confirms the persistent ultra-soft state of the source during this period (see Fig.~\ref{fig:lc-tt}).

The case of the 23 September 2016 observation (last VLBI trigger) is more complex. 
SRT could not participate in the session due to technical problems following a storm. The fringe detection threshold therefore became significantly larger, probably around 15--20 mJy. 
Cyg X-3 was not detected even though the calibrators were correctly observed during the session.  
A summary of the five VLBI observations is presented in Table~\ref{tab:vlbi-summary}.

\section{Discussion}

\subsection{X-ray and radio connections in August/September 2016}

Ejection processes are closely linked to accretion in X-ray binaries, but the mechanisms at the origin of the launch of relativistic jets are not well established \citep{Fender_2004, Markoff_2005}. 
%\citep[][and references therein]{Fender_2004, Markoff_2005}. 
%Energetic particles are accelerated away from the compact object and reach relativistic speeds in collimated jets.  They lose their energy via synchrotron radiation, inverse Compton emission and/or adiabatic losses, or via pion production in the case of baryonic jets \citep{Atoyan_1999}.

The hard X-ray light curve (15--50 keV) extracted from Swift/BAT indicated that the flux  of Cyg X-3 dramatically dropped starting from 15 August 2016 (MJD 57615) 
to reach an ultra-soft state ($\sim 0.01$ Crab flux) a few days later (see Fig.~\ref{fig:lc-tt}).
% It suddently increased on 29 August for three days ($\sim 0.05$ Crab flux), before decreasing again, in correspondence to the first VLBI observation at 22 GHz. 
The hardness ratio and X-ray light curves obtained with MAXI (2--4 keV) and Swift/BAT highlight changes of states in Cyg X-3, from the ultra-soft to the soft state during the giant flare (maximum of the peak on MJD 57650), 
%($\sim$ MJD $57645-57655$), 
then from the soft to the hard state at the end of the event (MJD 57660). 
This favors changes in the accretion rate, likely related to variations in the mass-loss from the dense stellar wind associated with the Wolf-Rayet companion star \citep{Kitamoto_1994, Watanabe_1994, Gies_2003}. 

The quenching of the jet in the ultra-soft X-ray state could be a direct consequence of high accretion rates,
with the inner radius of the disc approaching very close to the compact object \citep{Hjalmarsdotter_2009}.
Giant flares mark the end of this state and the transition to the soft state \citep{Corbel_2012}. In this way, they became predictable events \citep{Koljonen_2010, Trushkin_2016ATel9416}.
%are clearly connected to the transition from the ultra-soft state to the soft state
%The efficiency of the jet is direclty connected to the transition from the ultra-soft state to the soft state, which gives rise to giant flares \citep{Corbel_2012}. 

A correlation may be present between the duration of the ultra-soft state and the strength of the subsequent radio flare. The two last giant flares in March 2011 and September 2016 had an extensive radio and X-ray monitoring that covered the full duration of the events.  
The March 2011 flare followed a very long ultra-soft state of $\sim 36$ days, reached a peak flux density of $\sim 20$ Jy at 15 GHz, doubling the September 2016 flare flux density ($\sim 10$ Jy at 18.6 GHz) that occurred after a much shorter ultra-soft state ($\sim 23$ days). Even for the multiple flares in 
2006, the stronger double radio event in May peaking at $\sim$14 Jy at 15 GHz is preceeded by a longer ultra-soft state ($\sim 41$ days) with respect to the July event ($\sim$7 Jy at 11.2 GHz and $<$20 days ultra-soft state).

\subsection{A small flare precursor to the giant flare}

While Cyg X-3 was in the ultra-soft state, the hard X-ray flux suddenly increased for the three days following 29 August 2016 ($\sim 0.05$ Crab flux), before decreasing again on 1 September (see Fig.~\ref{fig:lc-tt}). 
Fast variations of the flux density were detected at 22 GHz during the first VLBI session triggered on the same day (MJD 57632.7). Figure~\ref{fig:lc-vlbi-15} shows the presence of two peaks of $\sim 440$ mJy and $\sim 250$ mJy with a duration of two hours each, while the off-peaks indicate a general diminution of the flux density. Images of the emitting component obtained during the first four hours of the observation highlighted an increase of the emission size from 0.6 to 0.9 mas (radius), as shown in Fig.\,\ref{fig:jet-1sept-size} and Fig.\,\ref{fig:plot-1sept-size}
at the moment of the first peak at $\sim 440$ mJy during the mini-flare. We measured an expansion of 0.3 mas in radius in 4 hours, equivalent to a velocity of 0.07--0.09$c$, assuming a distance to the source at 7--9 kpc. This velocity appears to be slower than the sound speed in mildly relativistic plasma and could be associated with the presence of a wind rather than a jet.
%This seems really too slow, it's slower than the sound speed in mildly relativistic plasma.   It doesn't make sense to me that a source as powerful as this one gives off such slow speeds unless you're seeing more of a wind than a jet.
%Extension:
%Radius 0.6 mas : $6-8\times 10^13$ cm, ($7-9$ kpc)
%Radius 0.9 mas : $9-12\times 10^13$ cm, ($7-9$ kpc)
%Velocity of expansion: 0.3 mas in 4 hours: $0.07-0.09c$.
%Our res
The source was not detected during the VLBI observation performed a few days later, on 3, 9, and 10 September (upper limit of 8 mJy at 22 GHz). %(MJD 57634.7).

The RATAN-600 light curve confirmed the presence of a mini-flare \citep{Trushkin_2016} that started on 30 August
%MJD 57630 
and lasted for $\sim 4$ days before Cyg X-3 came back to the quenched state ($\sim 30$ mJy at 11.2 GHz). The peak reached a maximum of $\sim$ 600--700 mJy at 8.2--11.2 GHz and $\sim 300$ mJy at 4.6 GHz on  1 September
%MJD 57632 
\citep{Trushkin_2016}. 
Our VLBI data obtained on 1 September 2016 correspond to the very beginning of the declining phase of the small flare, about 10 days before the onset of the rising phase of the giant flare.
%flux variations during the receding phase of the mini flare.
%We therefore observed the source at the very beginning and at the end of the receding phase of the small flare, about 10 days before the onset of the rising phase of the giant flare.
%The RATAN-600 light curve confirmed the presence of the small flare, with a peak maximum of roughly 0.7 Jy at $8.2-11.2$ GHz on 1 September \citep{Trushkin_2016}. 
%The flare lasted for $\sim 3$ days before Cyg X-3 came back to the quenched state ($\sim 30$ mJy at 11.2 GHz). 
The short-duration flare occurred close to the core, with an evidence of a very slight extension of the wind or jet.
%The short-duration flare occurred close to the core, without any evidence of an extended jet (jet not resolved, LIMIT ON JET SIZE ?).  
%Small flares with an amplitude $< 300$ mJy were detected with the VLBA at 15.3 GHz (Newell et al. 1998).
%%%\textit{The source core position is expected to change with time \citep{Miller-Jones_2004}.}
% small flares before the giant flare
Similar small radio flares have often been observed prior to giant flare events of Cyg X-3 \citep{Waltman_1994, Waltman_1995, Newell_1998, Kim_2013}. In particular, a VLBI observation performed in 2007 during a short-live flare (3 hours) with a flux density of 1.6 Jy gives similar results. The analysis of the Gaussian fits to the visibility amplitudes with time bin of 10\,min have shown an increase of the source size or a structural change during the mini-flare \citep{Kim_2013}. In both cases, we note that the emitting size slightly increases from the peak of the mini-flare until the end of the flare. As a consequence, the maximum of the flux density is reached before the small expansion of the wind or jet occured.
%The maximum of the flux intensity is reached most likely at the shock formation, giving birth to small blobs that then move sligthly from the core, but remains very close to the core.

%They may be explained by the dense stellar wind from the Wolf-Rayet compagnon star \citep{Kitamoto_1994, Watanabe_1994}.
%Heavy accretion, magnetic field lines twisted, critical point, then massive ejection of materiel into the jet (Waltman_1994). Variability of the radio emission: stellar wind density. During periods of high accretion, the process which supplies the radio jet during period of quiescent emission may have been quenched.
%and/or by jet ejection.
%%%%%%%%%%%%%\textit{Are they always precursors of giant flares?}
%Synchrotron self-absorption and free–free absorption by entrained thermal material play a larger role in determining the opacity than absorption in the stellar wind of the companion (Miller-Jones 2009).

\subsection{Spectral index evolution during the giant flare}

Single-dish observations of Cyg X-3 show evidence of flux density variations on a timescale of a few hours.
%, as illustrated in Figures~\ref{fig:single-dish} and \ref{fig:zoom-spec-ind}. 
The variations observed in the 7.2--8.5 GHz and 18.6--25.6 GHz light curves do not follow the same trend, in particular at the peak maximum, on 19 September 2016. While the flux density at lower frequencies remains quite stable, fast changes are observed at higher frequencies.
%particularly visible at high frequencies, from 18.6 to 25.6 GHz.
% and correspond to the peak of the giant flare.
The evolution of the spectral index testifies the high variability of the source.
%, in particular when the flare was at its maximum, on 19 September. 
The spectral index flattens ($\alpha \sim 0.3$) during the rise of the source flux, while it steepens ($\alpha \sim 0.6$) when the source flux declines at high frenquencies.
We note that the values of the spectral index are very similar to the ones associated with previous giant flares \citep[][and references therein]{Miller-Jones_2004} and with low-level flare events \citep{Miller-Jones_2009}. 
%Peaks of flux intensity correspond to $\alpha \sim 0.3-0.4$, whereas $\alpha \sim 0.7$ when the flux decreases.
However, the spectral steepening over a timescale of hours is for the first time highlighted. 
\citet{Miller-Jones_2004} studied several hypotheses based on energy-dependent loss mechanisms to explain the steepening on the scale of a few days. Synchrotron, bremsstrahlung, inverse Compton and leakage (diffusive escape) losses have been ruled out because of the too long timescales. The only mechanism capable of producing such a spectral evolution is related to light-travel time effects in plasmons \citep{van-der-Laan_1966}. Plasmons would evolve from optically thick to optically thin as they move outward from the core and expand.
A flat-spectrum core component dominates the flux density initially but fades over time. 
A steep spectral index would be reached when the plasmons become optically thin.
%%%This allows us to infer that the flux was decreasing during our first single-dish observations (MJD 57648.9) since $\alpha \sim 0.7$. The light curve obtained with the RATAN-600 confirms this statement, the flux density decreased sligthly on 17 September before reaching the prominent peak on 19 September \citep{Trushkin_2016}.

%ejection jets is probably optically thick due to synchrotron self-absorption or by thermal electrons mixed with relativistic ones
%We observed fast changes from $\alpha = 0.45\pm0.05$ to $0.28\pm0.05$ on less than 15 minutes. 
%$\alpha$ increased drastically in 2.5 hrs to reach $0.59\pm0.05$, and steepen again $0.78\pm0.05$
%Variations of the flux density were detected at 22 GHz on the scale of a few hours during the single-dish observation, as shown in  (peak flare). 
%On the same Fig.~\ref{fig:single-dish}, we provided the evolution of the spectral index $\alpha$, where $S_{\nu} \propto \nu^{-\alpha}$...
%Self-absorption by particles accelerated by central engine. Flat spectra: reacceleration of relativistic particles by shocks along the jet (model Marti et al. 1992, Waltman_1994).

\subsection{Comparison between the small and the giant flares}

Cyg X-3 was not detected during the last VLBI observation performed at 22 GHz, four days after the peak of the giant flare.
%($\sim 8$ Jy at 22.7 GHz) 
Medicina recorded a flux density of $3.1 \pm 0.1$ Jy at 8.5 GHz the same day,
%(MJD 57654.0).
while the RATAN-600 detected a radio emission of $\sim 2$ Jy at 21.7 GHz on 22 September 2016 and $\sim 0.8$ Jy on 24 September 2016 \citep{Trushkin_2016}. 
%This, however, remains much below the total flux density estimated from the RATAN-600 light curve, therefore it is likely that the source was strongly resolved out.
We can estimate the expected flux density at 22 GHz on 23 September, considering that
the fading of the flare follows an exponential law $\propto \mathrm{exp}[-(t-t_\mathrm{m})/2\mathrm{days}]$, where $t_\mathrm{m} = 57650.7$ the day of the peak maximum.
This results in a flux density estimated at $\sim1.4$ Jy, much higher than the upper limit we infered for this observation. The source was therefore strongly resolved out at mas scales.

In this way, our VLBI observations give us a direct comparison between the decay phases of the small/short-duration flare and the giant/longer flare in September 2016. 
%The fact that the source was not detected during the fading giant flare clearly indicates a different jet morphology and strength with respect to the small flare. 
Multi-frequency observations carried out with SRT on 21 September indicated that the spectrum of Cyg X-3 was optically thin two days after the peak of the flare. \citet{Trushkin_2016} confirm this statement and find a clear transition from optically thick to optically thin spectra right after the flare reached its highest brightness. Similar conclusions were drawn in the case of small flares \citep{Miller-Jones_2009}. Optically thick spectra are most likely attributed to synchrotron self-absorption or thermal electrons mixed with relativistic ones, whereas optically thin spectra are probably associated with ejecta in expansion moving outwards from the core \citep{Miller-Jones_2009}.
% expanding jets

The results we obtained with the single-dish and VLBI observations are consistent with the shock-in-jet scenario supported by \citet{Lindfors_2007}, \citet{Miller-Jones_2009} and \citet{Turler_2011}. These authors suggest that the differences in shape, amplitude, timescale and frequency range of the flares are related to the strength of the shocks along the jet.
Weaker and faster flares are produced closer to the core \citep{Turler_1999} whereas brighter flares evolve on longer time scales, peak at lower frequencies, and are the consequence of shocks forming further downstream in the jet. Shocks could provide a mechanism for the continuous replenishment of relativistic particles \citep{Atoyan_1999}.
However, shocks generally give an index of the power-law energy distribution of relativistic particles $\mathrm{p} > 2$ or $\alpha > 0.5$ with $\alpha = (\mathrm{p} -1)/2$. 
While it may be possible to produce a $\mathrm{p} < 2$ in shocks, 
magnetic reconnection in relativistic plasmas with a relatively high magnetization 
%($\sigma \geq 10$) 
could also explain the change from $\mathrm{p}=1.6$ to 2.2 or $\alpha=0.3$ to 0.6 \citep{Guo_2014, Sironi_2014, Sironi_2016}.
% Sironi, Giannios, & Petropoulou 2016 and references within).

%power-law energy distribution of cosmic rays accelerated in shocks

%The change from p=1.8 to 2.2 is actually quite difficult to explain from PIC simulations.  It may be possible to produce a p<2.0 in a shock, magnetic reconnection in a plasma with a relatively high magnetization (Sironi & Spitkovsky 2014; Guo, Li, Daughton & Liu 2014; Sironi, Giannios, & Petropoulou 2016 and references within).
%is magnetic reconnection the physical process generating the new population of very fast particles?
%I worry a bit about this, because generally a shock will give you p>2 or alpha > 0.5.   But you need something that goes from p=1.8 to 2.2, that is actually quite difficult, based on what we now know from PIC simulations.  While it may be possible to produce a p<2.0 in a shock, magnetic reconnection in a plasma with a relatively high magnetization (sigma >10) should also be mentioned (e.g. Sironi & Spitkovsky 2014; Guo, Li, Daughton & Liu 2014; Sironi, Giannios, & Petropoulou 2016 and references within).

% shape, amplitude, frequency range and timescale can be fairly described by varying only the strength of the shock and its build-up distance from the apex of the jet. A rapid build-up results in high-frequency outbursts evolving on short timescales, while slowly evolving, low-frequency outbursts form and evolve further out in the jet. 
%

The gamma-ray flares detected just before the mini radio flare with AGILE \citep{Piano_2016} and at the very onset of the giant radio flare with Fermi/LAT \citep{Cheung_2016} are also in agreement with our results and with the conclusions drawn from \citet{Corbel_2012} during the March 2011 giant flare. 
The gamma-ray activity is most likely related to shocks appearing at different distances along the jet. 
%These rapid variations in the relativistic jets imply changes in jet efficiency. 
%Shocks occuring closer to the core are consistent 
Particle acceleration (thanks to shocks or reconnection) happening closer to the core is consistent with
a brighter gamma-ray emission than that observed in shocks produced further downstream where the energy density in seed photon decreases, reducing inverse-Compton emission \citep{Dubus_2010}.
%A gamma-ray emission was detected during transitions in/out of the ultra-soft X-ray state. 

%besides the different intensity of the flares.
%found to be required in a plasmon model 
%the faster, weaker outbursts occur closer to the core, whereas the brighter outbursts occur further downstream. For a constant jet speed (as assumed in our modelling), this implies that shocks forming  further  downstream  take  longer  to  expand  by  a  given  factor, giving rise to a longer time-scale for the emission, and a lower peak frequency.
%with an increased electron normalization
%and magnetic field strength both playing a role in setting the strength of the outburst.
%peak at lower frequencies and evolve on longer time scales.
%Constraints on knot size?

\subsection{Comparison with previous giant flares}

In the following, we try to infer some constraints on the geometry and structure of the jet
%The non-detection VLBI of Cyg X-3 
associated with the fading giant flare on 23 September 2016, based on a comparison with previous giant flares of Cyg X-3.
%Cyg X-3 was not detected during the VLBI observation triggered on 23 September 2016 at 22 GHz, while the flux density was theoretically at $\sim 1.4$ Jy (see Section 3.2). The peak of the flare ($\sim 8$ Jy at 22.7 GHz) was reached four days prior to this observation. We note that the flux density maximum was similar to the 2001 flare. 
%allows us to infer some constraints on the geometry and structure of the jet.
Since the discovery of the first major radio flares by \citet{Gregory_1972}, 
%$\sim 50$ outbursts were detected with a peak radio flux density exceeding 1 Jy (2004 => need an update). Among them, 
Cyg X-3 has gone through a dozen of giant flare episodes exceeding 10 Jy \citep[][and reference therein]{Waltman_1995}. 
The last five giant flares occurred in February 1997 \citep{Mioduszewski_2001}, September 2001 \citep{Miller-Jones_2004}, May-July 2006 \citep{Pal_2009, Koljonen_2013}, March 2011 \citep{Corbel_2012} and the last one in September 2016 \citep{Trushkin_2016}. 

High-resolution images obtained with the VLBA clearly demonstrated the complex jet-like structures during the 1997 and 2001 flares.
A one-sided jet was detected in the south direction during the 1997 flare, with a speed $\geq 0.81c$ and a precession period $\geq 60$ days \citep{Mioduszewski_2001}. The jet emission extended over 50 mas two days after the peak of the flare (10 Jy at 15 GHz), and over 120 mas two days later. The corresponding synthesized beams were $\sim$ 3--5 mas while the proper motion of the jet was $> 20$ mas/day. This implies a movement by at least 2 beams during the $\sim 12$ hrs of the observations, which did not affect the images.
VLBA images obtained at 22 GHz during the peak maximum of the 2001 flare revealed a strong core emission \citep{Miller-Jones_2004}. The corresponding flux density was measured at 7.4 Jy with the VLA. A two-sided jet in an almost north-south orientation then appeared the following day, consisting of several discrete knots. A successive observation was triggered two days after the peak maximum and confirmed the expansion of the knot sequence with the fading flux density of $\sim 5$ Jy. The proper motions of individual knots, whose initial diameters are $\sim 8$ mas, were measured for the first time and  showed evidence of a 5-day jet precession period with a jet speed $\sim 0.63c$.

%The non-detection VLBI of Cyg X-3 during the fading giant flare on 23 September 2016 allows us to infer some constraints on the geometry and structure of the jet.
%Cyg X-3 was not detected during the VLBI observation triggered on 23 September 2016 at 22 GHz, while the flux density was theoretically at $\sim 1.4$ Jy (see Section 3.2). The peak of the flare ($\sim 8$ Jy at 22.7 GHz) was reached four days prior to this observation. We note that the flux density maximum was similar to the 2001 flare. 
%The angular resolution associated with our VLBI observation on 23 September 2016 was $\sim 1.$ mas (depending on the projected baseline in the \textit{uv}-plane).
%A two-sided jet was most likely emitted, with knots probably moving ballistically out from the core.
%The non-detection of the source can be explained by the knot size (growing from 8 to 16 mas in the case of the 2001 flare) and the precession period of the jet...
%Assuming a jet speed of $0.5c$
The angular scale associated with our VLBI observation on 23 September 2016
was $\sim 1$ mas, depending on the projected baseline in the
$(u,v)$-plane. In particular, we estimate a $5\sigma$ sensitivity of 20
mJy beam$^{-1}$, with a beam of 1.4 mas $\times$ 0.8 mas full-width at half
maximum (FWHM). This corresponds to a beam area of 0.88 mas$^2$. 
%The total flux density of the source estimated from the RATAN-600 light curve is 1.4 Jy. Our non detection therefore implies that the source was extended at least over 1400/20 beam areas, i.e. $\sim 60$ mas$^2$. If we assume a symmetric, two-sided ejection, we can determine the size of each region to be $\pi r^2 \ge 30$ mas$^2$. For a spherical blob, this implies a sphere radius of $r \ge 3$ mas. %, or about $10^{14}$ cm. 
The total flux density of the source estimated from the RATAN-600 light curve is 1.4 
Jy.  If we assume a symmetric, two-sided ejection, our non
detection therefore implies that each of the two $S=0.7$ Jy features is
distributed over a sky area $A$ such that $S/A \le 5\sigma$; i.e. $A \ge
(700 \ \mathrm{mJy})/(20 \ \mathrm{mJy beam}^{-1}) \sim 35 \ \mathrm{beam}
\sim 30 \ \mathrm{mas}^2$. For a circular component, we thus determine the
size of each region to be $\pi r^2 \ge 30$ mas$^2$, or $r \ge 3$ mas. 
If we further assume that the separation from the core is about $10\times$ faster than the blob
expansion, that implies a distance from the black hole of $\ge 30$ mas, or
3--4 $\times 10^{15}$ cm (67 AU), considering a location at 7--9 kpc.
%($\sim 9$ light-hr). 
Depending on when we assume the time of ejection, it is
straightforward to determine a lower limit to the projected jet knot velocity.
For example, assuming a blob formation at the peak radio emission as suggested by the change in the spectral index we observed and the clear transition from optically thick to optically thin jet \citep{Trushkin_2016} at that epoch (on 19 September), we would obtain a jet speed $> 0.3c$.
This is consistent with the plasmon expansion speed derived from previous giant flares. Instead, a later blob formation ($\le 1$ day from this VLBI observation) would imply a superluminal motion.

%This assumption is consistent with the change in the spectral index we detected at the pea

% Vicotria:
% I think this is something where you need to show number. Either  assume something reasonable for the ejection time and do the calculation or assume both the latest and the earliest ejection time that makes  sense and give numbers then. I know it all still would be just rough  estimates but I think they would be an interesting numbers to compare to for the reader.

\subsection{The peculiar case of Cyg X-3}

A comparison with well-known microquasars shows us that Cyg X-3 represents a unique and somehow very particular source for different reasons. 
%
% indicates a different behavior in outburst.
No other X-ray binaries have shown radio flux densities up to $\sim 20$ Jy during giant flare events. Major flares of 1--10 Jy have been observed in some transient black hole X-ray binaries, such as in SS433, V404 Cyg, GRO J1655$-$40 and A0620$-$00, all associated with hard-to-soft X-ray transitions. Based on spectral, velocity  and morphological characteristics, two types of radio jets have been identified during outbursts, corresponding to different X-ray states \citep{Fender_2009} and so different accretion regimes. \textit{A steady jet} appearing as a bright core \citep{Dhawan_2000}, with a flat or inverted spectrum indicating optically thick self-absorbed synchrotron emission is associated with the hard state \citep{Fender_2004}. 
%No core jet has ever been detected in the soft state. 
\textit{A transient jet}, corresponding to the ejection of optically thin radio plasmons moving away from the core of the system at relativistic speeds \citep{Mirabel_1994}, is instead associated with the transition between two types of intermediate states, which are themselves in between the hard and the soft state
%with the very high soft state 
\citep{Fender_2004}. Transitions from steady to transient jet occurs during outbursts, corresponding to an inversion of the spectrum from inverted to optically thin, as also seen in GRS 1915+105 and LS I $+61^\circ303$. 
% (Massi 2010). 

%GRO J1655-40 (Kalemci 2016)

In the case of Cyg X-3, giant flares occur at the end of the ultra-soft X-ray state \citep{Watanabe_1994} also defined as hyper-soft state by \citet{Koljonen_2010}, during the transition to a harder state. The jet is found to be optically thick during the rising phase of the giant flare, corresponding to the soft X-ray state, while the jet becomes optically thin at the peak and declining phase of the flare, which corresponds to soft-to-hard state transition \citep{Miller-Jones_2004, Trushkin_2016}. A similar transition from the steady to transient jet is therefore observed, however the association with X-ray states is clearly different from other transient X-ray binaries. 
The presence of strong stellar wind from the companion star could be at the origin of these differences.

GRO J1655$-$40 shows interesting similarities with Cyg X-3, in particular the presence of the hyper-soft X-ray state that could be associated with very high and unusual rate of accretion close to or above the Eddington limit \citep{Uttley_2015}, and strong radio flares up to 10 Jy. GRO J1655$-$40 is a black hole low-mass X-ray binary, that presents the  most powerful (possibly magnetically driven) disk wind among the other microquasars. 
%Hard X-ray flares have been detected during the transition from the soft to the ultra-soft  X-ray state, and also in the ultra-soft state. However, no giant flare has been observed after the 2005 ultra-soft X-ray state \citep{Kalemci_2016}.
Hard X-ray flares have been detected during the transition from the soft to the ultra-soft X-ray state, and also in the ultra-soft state. The first flare observed in 2005 during the transition from the soft to the ultra-soft state is coincident with an optically thin radio flare which is relatively weak compared to transitional radio flares observed from GRO J1655$-$40 in earlier outbursts
\citep{Kalemci_2016}. However, no giant flare has been observed after the
2005 ultra-soft X-ray state\footnote{\url{ www.aoc.nrao.edu/~mrupen/XRT/GRJ1655-40/grj1655-40.shtml}}.

\section{Conclusions}

Cyg X-3 represents an exceptional target that offers the possibility to better understand 
the relationship between accretion state and jet launching mechanisms, 
%I would say more than this, it's really giving a chance 
and to compare the effect of accretion geometry (companion star, orbital parameters) to other sources with more typical behaviour,
in particular in the extreme cases of giant flares. These very bright and spectacular radio events are rare, clearly associated with the ultra-soft X-ray state and gamma-ray emission.
After 5.5 years of quiescence, a giant flare occurred in September 2016. 
Single-dish observations performed with Medicina and SRT followed the evolution of the peak over 6 days in six frequency ranges. The observed frequencies are complementary with the ones used in the RATAN-600. Moreover, the long exposures provided with the Italian radio telescopes allowed us to infer variation of the radio emission on short timescale. In particular, we highlighted a decline of the flux density at high radio frequency with a steepening of the spectrum from $\nu^{-0.3}$ to $\nu^{-0.6}$ within $\sim 5$ hours at the peak of the flare. It is the first time that such a steepening is observed on the hour scale, which gives support to plasmon evolution from optically thick to optically thin as they move outward from the core and expand. 

VLBI observations were triggered at 22 GHz at different phases of the 2016 flare episode.
Flux variations were detected within 2 hours ten days before the onset of the giant radio flare.  They are associated with the declining phase of a mini and short-lived flare produced close to the core. We measured a slight increase of the source size at the moment of the highest peak of the mini-flare.
A VLBI observation performed 4 days after the peak of the giant flare allows us to infer constraints on the size and velocity of the jet. The jet emission was most likely extended over 30 mas with a jet knot velocity $> 0.3c$ assuming a blob formation at the peak emission as suggested by the change in the spectral index we observed.

The complementarity between single-dish and VLBI observations is essential in order to better understand ejection mechanisms during giant flare episodes.
The data recorded with the recently commissioned SRT confirm its excellent capabilities operating as single-dish and VLBI antenna. The selection of a few EVN telescopes as a EVN lite is very useful to provide VLBI observations of such rare events. For the next giant flares, it would be interesting to perform multi-frequency single-dish observations in parallel to VLBI observations from the peak of the flare in order to better track the plasmon evolution from optically thick to optically thin, and directly see the link between changes in the spectral index and jet morphology
%understand the change from optically thick to optically thin jets 
on relative short time scales (a few hours). Moreover, it would be more fitted to consider EVN-lite observations at lower frequency during the decay phase of the flare in order to study the morphology, velocity and evolution of the jet.

%It might be nice to say a few things about what you would do differently for the next giant flare?   Ie.. you say these are now understood enough to trigger on, so how would the results in this paper impact your decisions for the next giant flare, in terms of what you would try to capture, what questions you could answer?   

%Emrah
%the variability you have observed in spectral index variations is shorter than previous work? You highlighted that in the abstract, and if this is the case, it is the point that you need to build the scientific case upon. You have talked about the plasmons, but maybe you could say this work strengthens this possibility among others?

\section*{Acknowledgements}

The authors would like to thank S. Trushkin for the useful discussions about Cyg X-3.
M.P. acknowledges financial support from the RAS (CRP-25476).
%M.P. acknowledges the financial support from the RAS through grant CRP-25476.
S.C. acknowledges the financial support from the UnivEarthS Labex program of Sorbonne Paris Cit\'{e} (ANR-10-LABX-0023 and ANR-11- IDEX-0005-02). 
E.K. acknowledges support from TUBITAK Grant 115F488.
The Sardinia Radio Telescope is funded by the Department of University and Research (MIUR), the Italian Space Agency (ASI), and the Autonomous Region of Sardinia (RAS), and is operated as a National Facility by the National Institute for Astrophysics (INAF).
Based on observations with the Medicina telescope operated by INAF - Istituto di Radioastronomia.

%%%%%%%%%%%%%%%%%%%%%%%%%%%%%%%%%%%%%%%%%%%%%%%%%%



\bibliographystyle{mnras}
\bibliography{biblio.bib} % if your bibtex file is called example.bib

\begin{thebibliography}{}
\makeatletter
\relax
\def\mn@urlcharsother{\let\do\@makeother \do\$\do\&\do\#\do\^\do\_\do\%\do\~}
\def\mn@doi{\begingroup\mn@urlcharsother \@ifnextchar [ {\mn@doi@}
  {\mn@doi@[]}}
\def\mn@doi@[#1]#2{\def\@tempa{#1}\ifx\@tempa\@empty \href
  {http://dx.doi.org/#2} {doi:#2}\else \href {http://dx.doi.org/#2} {#1}\fi
  \endgroup}
\def\mn@eprint#1#2{\mn@eprint@#1:#2::\@nil}
\def\mn@eprint@arXiv#1{\href {http://arxiv.org/abs/#1} {{\tt arXiv:#1}}}
\def\mn@eprint@dblp#1{\href {http://dblp.uni-trier.de/rec/bibtex/#1.xml}
  {dblp:#1}}
\def\mn@eprint@#1:#2:#3:#4\@nil{\def\@tempa {#1}\def\@tempb {#2}\def\@tempc
  {#3}\ifx \@tempc \@empty \let \@tempc \@tempb \let \@tempb \@tempa \fi \ifx
  \@tempb \@empty \def\@tempb {arXiv}\fi \@ifundefined
  {mn@eprint@\@tempb}{\@tempb:\@tempc}{\expandafter \expandafter \csname
  mn@eprint@\@tempb\endcsname \expandafter{\@tempc}}}

\bibitem[\protect\citeauthoryear{{Atoyan} \& {Aharonian}}{{Atoyan} \&
  {Aharonian}}{1999}]{Atoyan_1999}
{Atoyan} A.~M.,  {Aharonian} F.~A.,  1999, \mn@doi [\mnras]
  {10.1046/j.1365-8711.1999.02172.x}, \href
  {http://cdsads.u-strasbg.fr/abs/1999MNRAS.302..253A} {302, 253}

\bibitem[\protect\citeauthoryear{{Belloni}}{{Belloni}}{2010}]{Belloni_2010}
{Belloni} T.~M.,  2010, in {Belloni} T.,  ed.,  Lecture Notes in Physics,
  Berlin Springer Verlag Vol. 794, Lecture Notes in Physics, Berlin Springer
  Verlag. p.~53 (\mn@eprint {arXiv} {0909.2474}),
  \mn@doi{10.1007/978-3-540-76937-8_3}

\bibitem[\protect\citeauthoryear{{Bolli} et~al.,}{{Bolli}
  et~al.}{2015}]{Bolli_2015}
{Bolli} P.,  et~al., 2015, \mn@doi [Journal of Astronomical Instrumentation]
  {10.1142/S2251171715500087}, \href
  {http://cdsads.u-strasbg.fr/abs/2015JAI.....450008B} {4, 1550008}

\bibitem[\protect\citeauthoryear{{Bonnet-Bidaud} \& {Chardin}}{{Bonnet-Bidaud}
  \& {Chardin}}{1988}]{Bonnet-Bidaud_1988}
{Bonnet-Bidaud} J.~M.,  {Chardin} G.,  1988, \mn@doi [\physrep]
  {10.1016/0370-1573(88)90083-X}, \href
  {http://cdsads.u-strasbg.fr/abs/1988PhR...170..325B} {170, 325}

\bibitem[\protect\citeauthoryear{{Cheung} \& {Loh}}{{Cheung} \&
  {Loh}}{2016}]{Cheung_2016}
{Cheung} C.~C.,  {Loh} A.,  2016, The Astronomer's Telegram, \href
  {http://cdsads.u-strasbg.fr/abs/2016ATel.9502....1C} {9502}

\bibitem[\protect\citeauthoryear{{Corbel} et~al.,}{{Corbel}
  et~al.}{2012}]{Corbel_2012}
{Corbel} S.,  et~al., 2012, \mn@doi [\mnras]
  {10.1111/j.1365-2966.2012.20517.x}, \href
  {http://cdsads.u-strasbg.fr/abs/2012MNRAS.421.2947C} {421, 2947}

\bibitem[\protect\citeauthoryear{{Deller} et~al.,}{{Deller}
  et~al.}{2011}]{Deller_2011}
{Deller} A.~T.,  et~al., 2011, \mn@doi [\pasp] {10.1086/658907}, \href
  {http://cdsads.u-strasbg.fr/abs/2011PASP..123..275D} {123, 275}

\bibitem[\protect\citeauthoryear{{Dhawan}, {Mirabel}  \&
  {Rodr{\'{\i}}guez}}{{Dhawan} et~al.}{2000}]{Dhawan_2000}
{Dhawan} V.,  {Mirabel} I.~F.,   {Rodr{\'{\i}}guez} L.~F.,  2000, \mn@doi
  [\apj] {10.1086/317088}, \href
  {http://cdsads.u-strasbg.fr/abs/2000ApJ...543..373D} {543, 373}

\bibitem[\protect\citeauthoryear{{Dubus}, {Cerutti}  \& {Henri}}{{Dubus}
  et~al.}{2010}]{Dubus_2010}
{Dubus} G.,  {Cerutti} B.,   {Henri} G.,  2010, \mn@doi [\mnras]
  {10.1111/j.1745-3933.2010.00834.x}, \href
  {http://cdsads.u-strasbg.fr/abs/2010MNRAS.404L..55D} {404, L55}

\bibitem[\protect\citeauthoryear{{Egron}, {Pellizzoni}, {Iacolina}, {Loru},
  {Righini}, {Trois}  \& {SRT Astrophysical Validation Team}}{{Egron}
  et~al.}{2016a}]{Egron_2016a}
{Egron} E.,  {Pellizzoni} A.,  {Iacolina} M.~N.,  {Loru} S.,  {Righini} S.,
  {Trois} A.,   {SRT Astrophysical Validation Team} 2016a, INAF - Osservatorio
  Astronomico di Cagliari. Internal Report N.59, \href
  {http://www.oa-cagliari.inaf.it/area.php?page_id=10} {}

\bibitem[\protect\citeauthoryear{{Egron} et~al.,}{{Egron}
  et~al.}{2016b}]{Egron_2016b}
{Egron} E.,  et~al., 2016b, The Astronomer's Telegram, \href
  {http://cdsads.u-strasbg.fr/abs/2016ATel.9508....1E} {9508}

\bibitem[\protect\citeauthoryear{{Fender}, {Hanson}  \& {Pooley}}{{Fender}
  et~al.}{1999}]{Fender_1999}
{Fender} R.~P.,  {Hanson} M.~M.,   {Pooley} G.~G.,  1999, \mn@doi [\mnras]
  {10.1046/j.1365-8711.1999.02726.x}, \href
  {http://cdsads.u-strasbg.fr/abs/1999MNRAS.308..473F} {308, 473}

\bibitem[\protect\citeauthoryear{{Fender}, {Belloni}  \& {Gallo}}{{Fender}
  et~al.}{2004}]{Fender_2004}
{Fender} R.~P.,  {Belloni} T.~M.,   {Gallo} E.,  2004, \mn@doi [\mnras]
  {10.1111/j.1365-2966.2004.08384.x}, \href
  {http://cdsads.u-strasbg.fr/abs/2004MNRAS.355.1105F} {355, 1105}

\bibitem[\protect\citeauthoryear{{Fender}, {Homan}  \& {Belloni}}{{Fender}
  et~al.}{2009}]{Fender_2009}
{Fender} R.~P.,  {Homan} J.,   {Belloni} T.~M.,  2009, \mn@doi [\mnras]
  {10.1111/j.1365-2966.2009.14841.x}, \href
  {http://cdsads.u-strasbg.fr/abs/2009MNRAS.396.1370F} {396, 1370}

\bibitem[\protect\citeauthoryear{{Fermi LAT Collaboration} et~al.,}{{Fermi LAT
  Collaboration} et~al.}{2009}]{Fermi_2009}
{Fermi LAT Collaboration} et~al., 2009, \mn@doi [Science]
  {10.1126/science.1182174}, \href
  {http://cdsads.u-strasbg.fr/abs/2009Sci...326.1512F} {326, 1512}

\bibitem[\protect\citeauthoryear{{Gallo}, {Fender}  \& {Pooley}}{{Gallo}
  et~al.}{2003}]{Gallo_2003}
{Gallo} E.,  {Fender} R.~P.,   {Pooley} G.~G.,  2003, \mn@doi [\mnras]
  {10.1046/j.1365-8711.2003.06791.x}, \href
  {http://cdsads.u-strasbg.fr/abs/2003MNRAS.344...60G} {344, 60}

\bibitem[\protect\citeauthoryear{{Giacconi}, {Gorenstein}, {Gursky}  \&
  {Waters}}{{Giacconi} et~al.}{1967}]{Giacconi_1967}
{Giacconi} R.,  {Gorenstein} P.,  {Gursky} H.,   {Waters} J.~R.,  1967, \mn@doi
  [\apjl] {10.1086/180028}, \href
  {http://cdsads.u-strasbg.fr/abs/1967ApJ...148L.119G} {148, L119}

\bibitem[\protect\citeauthoryear{{Gies} et~al.,}{{Gies}
  et~al.}{2003}]{Gies_2003}
{Gies} D.~R.,  et~al., 2003, \mn@doi [\apj] {10.1086/345345}, \href
  {http://cdsads.u-strasbg.fr/abs/2003ApJ...583..424G} {583, 424}

\bibitem[\protect\citeauthoryear{{Gregory} \& {Kronberg}}{{Gregory} \&
  {Kronberg}}{1972}]{Gregory_1972}
{Gregory} P.~C.,  {Kronberg} P.~P.,  1972, \mn@doi [\nat] {10.1038/239440a0},
  \href {http://cdsads.u-strasbg.fr/abs/1972Natur.239..440G} {239, 440}

\bibitem[\protect\citeauthoryear{{Greisen}}{{Greisen}}{2003}]{Greisen_2003}
{Greisen} E.~W.,  2003, \mn@doi [Information Handling in Astronomy - Historical
  Vistas] {10.1007/0-306-48080-8_7}, \href
  {http://cdsads.u-strasbg.fr/abs/2003ASSL..285..109G} {285, 109}

\bibitem[\protect\citeauthoryear{{Guo}, {Li}, {Daughton}  \& {Liu}}{{Guo}
  et~al.}{2014}]{Guo_2014}
{Guo} F.,  {Li} H.,  {Daughton} W.,   {Liu} Y.-H.,  2014, \mn@doi [Physical
  Review Letters] {10.1103/PhysRevLett.113.155005}, \href
  {http://cdsads.u-strasbg.fr/abs/2014PhRvL.113o5005G} {113, 155005}

\bibitem[\protect\citeauthoryear{{Hjalmarsdotter}, {Zdziarski}, {Larsson},
  {Beckmann}, {McCollough}, {Hannikainen}  \& {Vilhu}}{{Hjalmarsdotter}
  et~al.}{2008}]{Hjalmarsdotter_2008}
{Hjalmarsdotter} L.,  {Zdziarski} A.~A.,  {Larsson} S.,  {Beckmann} V.,
  {McCollough} M.,  {Hannikainen} D.~C.,   {Vilhu} O.,  2008, \mn@doi [\mnras]
  {10.1111/j.1365-2966.2007.12688.x}, \href
  {http://cdsads.u-strasbg.fr/abs/2008MNRAS.384..278H} {384, 278}

\bibitem[\protect\citeauthoryear{{Hjalmarsdotter}, {Zdziarski}, {Szostek}  \&
  {Hannikainen}}{{Hjalmarsdotter} et~al.}{2009}]{Hjalmarsdotter_2009}
{Hjalmarsdotter} L.,  {Zdziarski} A.~A.,  {Szostek} A.,   {Hannikainen} D.~C.,
  2009, \mn@doi [\mnras] {10.1111/j.1365-2966.2008.14036.x}, \href
  {http://cdsads.u-strasbg.fr/abs/2009MNRAS.392..251H} {392, 251}

\bibitem[\protect\citeauthoryear{{Kalemci}, {Begelman}, {Maccarone}, {Din{\c
  c}er}, {Russell}, {Bailyn}  \& {Tomsick}}{{Kalemci}
  et~al.}{2016}]{Kalemci_2016}
{Kalemci} E.,  {Begelman} M.~C.,  {Maccarone} T.~J.,  {Din{\c c}er} T.,
  {Russell} T.~D.,  {Bailyn} C.,   {Tomsick} J.~A.,  2016, \mn@doi [\mnras]
  {10.1093/mnras/stw2002}, \href
  {http://cdsads.u-strasbg.fr/abs/2016MNRAS.463..615K} {463, 615}

\bibitem[\protect\citeauthoryear{{Kim}, {Kim}, {Kurayama}, {Honma}, {Sasao}  \&
  {Kim}}{{Kim} et~al.}{2013}]{Kim_2013}
{Kim} J.-S.,  {Kim} S.-W.,  {Kurayama} T.,  {Honma} M.,  {Sasao} T.,   {Kim}
  S.~J.,  2013, \mn@doi [\apj] {10.1088/0004-637X/772/1/41}, \href
  {http://cdsads.u-strasbg.fr/abs/2013ApJ...772...41K} {772, 41}

\bibitem[\protect\citeauthoryear{{Kitamoto}, {Miyamoto}, {Waltman}, {Fiedler},
  {Johnston}  \& {Ghigo}}{{Kitamoto} et~al.}{1994}]{Kitamoto_1994}
{Kitamoto} S.,  {Miyamoto} S.,  {Waltman} E.~B.,  {Fiedler} R.~L.,  {Johnston}
  K.~J.,   {Ghigo} F.~D.,  1994, \aap, \href
  {http://cdsads.u-strasbg.fr/abs/1994A%26A...281L..85K} {281, L85}

\bibitem[\protect\citeauthoryear{{Koch-Miramond}, {{\'A}brah{\'a}m}, {Fuchs},
  {Bonnet-Bidaud}  \& {Claret}}{{Koch-Miramond}
  et~al.}{2002}]{Koch-Miramond_2002}
{Koch-Miramond} L.,  {{\'A}brah{\'a}m} P.,  {Fuchs} Y.,  {Bonnet-Bidaud} J.-M.,
    {Claret} A.,  2002, \mn@doi [\aap] {10.1051/0004-6361:20021273}, \href
  {http://cdsads.u-strasbg.fr/abs/2002A%26A...396..877K} {396, 877}

\bibitem[\protect\citeauthoryear{{Koljonen}, {Hannikainen}, {McCollough},
  {Pooley}  \& {Trushkin}}{{Koljonen} et~al.}{2010}]{Koljonen_2010}
{Koljonen} K.~I.~I.,  {Hannikainen} D.~C.,  {McCollough} M.~L.,  {Pooley}
  G.~G.,   {Trushkin} S.~A.,  2010, \mn@doi [\mnras]
  {10.1111/j.1365-2966.2010.16722.x}, \href
  {http://cdsads.u-strasbg.fr/abs/2010MNRAS.406..307K} {406, 307}

\bibitem[\protect\citeauthoryear{{Koljonen}, {McCollough}, {Hannikainen}  \&
  {Droulans}}{{Koljonen} et~al.}{2013}]{Koljonen_2013}
{Koljonen} K.~I.~I.,  {McCollough} M.~L.,  {Hannikainen} D.~C.,   {Droulans}
  R.,  2013, \mn@doi [\mnras] {10.1093/mnras/sts404}, \href
  {http://cdsads.u-strasbg.fr/abs/2013MNRAS.429.1173K} {429, 1173}

\bibitem[\protect\citeauthoryear{{Krimm} et~al.,}{{Krimm}
  et~al.}{2013}]{Krimm_2013}
{Krimm} H.~A.,  et~al., 2013, \mn@doi [\apjs] {10.1088/0067-0049/209/1/14},
  \href {http://cdsads.u-strasbg.fr/abs/2013ApJS..209...14K} {209, 14}

\bibitem[\protect\citeauthoryear{{Lindfors}, {T{\"u}rler}, {Hannikainen},
  {Pooley}, {Tammi}, {Trushkin}  \& {Valtaoja}}{{Lindfors}
  et~al.}{2007}]{Lindfors_2007}
{Lindfors} E.~J.,  {T{\"u}rler} M.,  {Hannikainen} D.~C.,  {Pooley} G.,
  {Tammi} J.,  {Trushkin} S.~A.,   {Valtaoja} E.,  2007, \mn@doi [\aap]
  {10.1051/0004-6361:20077620}, \href
  {http://cdsads.u-strasbg.fr/abs/2007A%26A...473..923L} {473, 923}

\bibitem[\protect\citeauthoryear{{Ling}, {Zhang}  \& {Tang}}{{Ling}
  et~al.}{2009}]{Ling_2009}
{Ling} Z.,  {Zhang} S.~N.,   {Tang} S.,  2009, \mn@doi [\apj]
  {10.1088/0004-637X/695/2/1111}, \href
  {http://cdsads.u-strasbg.fr/abs/2009ApJ...695.1111L} {695, 1111}

\bibitem[\protect\citeauthoryear{{Markoff}, {Nowak}  \& {Wilms}}{{Markoff}
  et~al.}{2005}]{Markoff_2005}
{Markoff} S.,  {Nowak} M.~A.,   {Wilms} J.,  2005, \mn@doi [\apj]
  {10.1086/497628}, \href {http://cdsads.u-strasbg.fr/abs/2005ApJ...635.1203M}
  {635, 1203}

\bibitem[\protect\citeauthoryear{{Mart{\'{\i}}}, {Paredes}  \&
  {Peracaula}}{{Mart{\'{\i}}} et~al.}{2001}]{Marti_2001}
{Mart{\'{\i}}} J.,  {Paredes} J.~M.,   {Peracaula} M.,  2001, \mn@doi [\aap]
  {10.1051/0004-6361:20010907}, \href
  {http://cdsads.u-strasbg.fr/abs/2001A%26A...375..476M} {375, 476}

\bibitem[\protect\citeauthoryear{{McClintock} \& {Remillard}}{{McClintock} \&
  {Remillard}}{2006}]{McClintock_2006}
{McClintock} J.~E.,  {Remillard} R.~A.,  2006, {Black hole binaries}.
pp 157--213

\bibitem[\protect\citeauthoryear{{McCollough} et~al.,}{{McCollough}
  et~al.}{1999}]{McCollough_1999}
{McCollough} M.~L.,  et~al., 1999, \mn@doi [\apj] {10.1086/307241}, \href
  {http://cdsads.u-strasbg.fr/abs/1999ApJ...517..951M} {517, 951}

\bibitem[\protect\citeauthoryear{{McCollough}, {Corrales}  \&
  {Dunham}}{{McCollough} et~al.}{2016}]{McCollough_2016}
{McCollough} M.~L.,  {Corrales} L.,   {Dunham} M.~M.,  2016, \mn@doi [\apjl]
  {10.3847/2041-8205/830/2/L36}, \href
  {http://cdsads.u-strasbg.fr/abs/2016ApJ...830L..36M} {830, L36}

\bibitem[\protect\citeauthoryear{{Miller-Jones}, {Blundell}, {Rupen},
  {Mioduszewski}, {Duffy}  \& {Beasley}}{{Miller-Jones}
  et~al.}{2004}]{Miller-Jones_2004}
{Miller-Jones} J.~C.~A.,  {Blundell} K.~M.,  {Rupen} M.~P.,  {Mioduszewski}
  A.~J.,  {Duffy} P.,   {Beasley} A.~J.,  2004, \mn@doi [\apj]
  {10.1086/379706}, \href {http://cdsads.u-strasbg.fr/abs/2004ApJ...600..368M}
  {600, 368}

\bibitem[\protect\citeauthoryear{{Miller-Jones}, {Rupen}, {T{\"u}rler},
  {Lindfors}, {Blundell}  \& {Pooley}}{{Miller-Jones}
  et~al.}{2009}]{Miller-Jones_2009}
{Miller-Jones} J.~C.~A.,  {Rupen} M.~P.,  {T{\"u}rler} M.,  {Lindfors} E.~J.,
  {Blundell} K.~M.,   {Pooley} G.~G.,  2009, \mn@doi [\mnras]
  {10.1111/j.1365-2966.2008.14279.x}, \href
  {http://cdsads.u-strasbg.fr/abs/2009MNRAS.394..309M} {394, 309}

\bibitem[\protect\citeauthoryear{{Mioduszewski}, {Rupen}, {Hjellming}, {Pooley}
   \& {Waltman}}{{Mioduszewski} et~al.}{2001}]{Mioduszewski_2001}
{Mioduszewski} A.~J.,  {Rupen} M.~P.,  {Hjellming} R.~M.,  {Pooley} G.~G.,
  {Waltman} E.~B.,  2001, \mn@doi [\apj] {10.1086/320965}, \href
  {http://cdsads.u-strasbg.fr/abs/2001ApJ...553..766M} {553, 766}

\bibitem[\protect\citeauthoryear{{Mirabel} \& {Rodr{\'{\i}}guez}}{{Mirabel} \&
  {Rodr{\'{\i}}guez}}{1994}]{Mirabel_1994}
{Mirabel} I.~F.,  {Rodr{\'{\i}}guez} L.~F.,  1994, \mn@doi [\nat]
  {10.1038/371046a0}, \href
  {http://cdsads.u-strasbg.fr/abs/1994Natur.371...46M} {371, 46}

\bibitem[\protect\citeauthoryear{{Mirabel} \& {Rodr{\'{\i}}guez}}{{Mirabel} \&
  {Rodr{\'{\i}}guez}}{1999}]{Mirabel_1999}
{Mirabel} I.~F.,  {Rodr{\'{\i}}guez} L.~F.,  1999, \mn@doi [\araa]
  {10.1146/annurev.astro.37.1.409}, \href
  {http://cdsads.u-strasbg.fr/abs/1999ARA%26A..37..409M} {37, 409}

\bibitem[\protect\citeauthoryear{{Newell}, {Garrett}  \& {Spencer}}{{Newell}
  et~al.}{1998}]{Newell_1998}
{Newell} S.~J.,  {Garrett} M.~A.,   {Spencer} R.~E.,  1998, \mn@doi [\mnras]
  {10.1046/j.1365-8711.1998.01230.x}, \href
  {http://cdsads.u-strasbg.fr/abs/1998MNRAS.293L..17N} {293, L17}

\bibitem[\protect\citeauthoryear{{Orlati}, {Bartolini}, {Buttu}, {Fara},
  {Migoni}, {Poppi}  \& {Righini}}{{Orlati} et~al.}{2016}]{Orlati_2016}
{Orlati} A.,  {Bartolini} M.,  {Buttu} M.,  {Fara} A.,  {Migoni} C.,  {Poppi}
  S.,   {Righini} S.,  2016, in Society of Photo-Optical Instrumentation
  Engineers (SPIE) Conference Series. p. 991310, \mn@doi{10.1117/12.2232581}

\bibitem[\protect\citeauthoryear{{Ott}, {Witzel}, {Quirrenbach}, {Krichbaum},
  {Standke}, {Schalinski}  \& {Hummel}}{{Ott} et~al.}{1994}]{Ott_1994}
{Ott} M.,  {Witzel} A.,  {Quirrenbach} A.,  {Krichbaum} T.~P.,  {Standke}
  K.~J.,  {Schalinski} C.~J.,   {Hummel} C.~A.,  1994, \aap, \href
  {http://cdsads.u-strasbg.fr/abs/1994A%26A...284..331O} {284, 331}

\bibitem[\protect\citeauthoryear{{Pal}, {Ishwara-Chandra}  \& {Rao}}{{Pal}
  et~al.}{2009}]{Pal_2009}
{Pal} S.,  {Ishwara-Chandra} C.~H.,   {Rao} A.~P.,  2009, in {Saikia} D.~J.,
  {Green} D.~A.,  {Gupta} Y.,   {Venturi} T.,  eds,  Astronomical Society of
  the Pacific Conference Series Vol. 407, The Low-Frequency Radio Universe.
  p.~277

\bibitem[\protect\citeauthoryear{{Parsignault} et~al.,}{{Parsignault}
  et~al.}{1972}]{Parsignault_1972}
{Parsignault} D.~R.,  et~al., 1972, \mn@doi [Nature Physical Science]
  {10.1038/physci239123a0}, \href
  {http://cdsads.u-strasbg.fr/abs/1972NPhS..239..123P} {239, 123}

\bibitem[\protect\citeauthoryear{{Perley} \& {Butler}}{{Perley} \&
  {Butler}}{2013}]{Perley_2013}
{Perley} R.~A.,  {Butler} B.~J.,  2013, \mn@doi [\apjs]
  {10.1088/0067-0049/204/2/19}, \href
  {http://cdsads.u-strasbg.fr/abs/2013ApJS..204...19P} {204, 19}

\bibitem[\protect\citeauthoryear{{Piano} et~al.,}{{Piano}
  et~al.}{2016}]{Piano_2016}
{Piano} G.,  et~al., 2016, The Astronomer's Telegram, \href
  {http://cdsads.u-strasbg.fr/abs/2016ATel.9429....1P} {9429}

\bibitem[\protect\citeauthoryear{{Prandoni} et~al.,}{{Prandoni}
  et~al.}{2017}]{Prandoni_2017}
{Prandoni} I.,  et~al., 2017, ArXiv e-prints, accepted for publication in A\&A,
  \href {http://cdsads.u-strasbg.fr/abs/2017arXiv170309673P} {}

\bibitem[\protect\citeauthoryear{{Predehl}, {Burwitz}, {Paerels}  \&
  {Tr{\"u}mper}}{{Predehl} et~al.}{2000}]{Predehl_2000}
{Predehl} P.,  {Burwitz} V.,  {Paerels} F.,   {Tr{\"u}mper} J.,  2000, \aap,
  \href {http://cdsads.u-strasbg.fr/abs/2000A%26A...357L..25P} {357, L25}

\bibitem[\protect\citeauthoryear{{Shrader}, {Titarchuk}  \&
  {Shaposhnikov}}{{Shrader} et~al.}{2010}]{Shrader_2010}
{Shrader} C.~R.,  {Titarchuk} L.,   {Shaposhnikov} N.,  2010, \mn@doi [\apj]
  {10.1088/0004-637X/718/1/488}, \href
  {http://cdsads.u-strasbg.fr/abs/2010ApJ...718..488S} {718, 488}

\bibitem[\protect\citeauthoryear{{Sironi} \& {Spitkovsky}}{{Sironi} \&
  {Spitkovsky}}{2014}]{Sironi_2014}
{Sironi} L.,  {Spitkovsky} A.,  2014, \mn@doi [\apjl]
  {10.1088/2041-8205/783/1/L21}, \href
  {http://cdsads.u-strasbg.fr/abs/2014ApJ...783L..21S} {783, L21}

\bibitem[\protect\citeauthoryear{{Sironi}, {Giannios}  \&
  {Petropoulou}}{{Sironi} et~al.}{2016}]{Sironi_2016}
{Sironi} L.,  {Giannios} D.,   {Petropoulou} M.,  2016, \mn@doi [\mnras]
  {10.1093/mnras/stw1620}, \href
  {http://cdsads.u-strasbg.fr/abs/2016MNRAS.462...48S} {462, 48}

\bibitem[\protect\citeauthoryear{{Szostek} \& {Zdziarski}}{{Szostek} \&
  {Zdziarski}}{2004}]{Szostek_2004}
{Szostek} A.,  {Zdziarski} A.~A.,  2004, ArXiv Astrophysics e-prints, \href
  {http://cdsads.u-strasbg.fr/abs/2004astro.ph..1265S} {}

\bibitem[\protect\citeauthoryear{{Szostek} \& {Zdziarski}}{{Szostek} \&
  {Zdziarski}}{2008}]{Szostek_2008}
{Szostek} A.,  {Zdziarski} A.~A.,  2008, \mn@doi [\mnras]
  {10.1111/j.1365-2966.2008.13073.x}, \href
  {http://cdsads.u-strasbg.fr/abs/2008MNRAS.386..593S} {386, 593}

\bibitem[\protect\citeauthoryear{{Szostek}, {Zdziarski}  \&
  {McCollough}}{{Szostek} et~al.}{2008}]{Szostek_2008b}
{Szostek} A.,  {Zdziarski} A.~A.,   {McCollough} M.~L.,  2008, \mn@doi [\mnras]
  {10.1111/j.1365-2966.2008.13479.x}, \href
  {http://cdsads.u-strasbg.fr/abs/2008MNRAS.388.1001S} {388, 1001}

\bibitem[\protect\citeauthoryear{{Tavani} et~al.,}{{Tavani}
  et~al.}{2009}]{Tavani_2009}
{Tavani} M.,  et~al., 2009, \mn@doi [\nat] {10.1038/nature08578}, \href
  {http://cdsads.u-strasbg.fr/abs/2009Natur.462..620T} {462, 620}

\bibitem[\protect\citeauthoryear{{Trushkin}, {Nizhelskij}, {Tsybulev}  \&
  {Zhekanis}}{{Trushkin} et~al.}{2016a}]{Trushkin_2016}
{Trushkin} S.~A.,  {Nizhelskij} N.~A.,  {Tsybulev} P.~G.,   {Zhekanis} G.~V.,
  2016a, preprint, \href {http://cdsads.u-strasbg.fr/abs/2016arXiv161200634T}
  {} (\mn@eprint {arXiv} {1612.00634})

\bibitem[\protect\citeauthoryear{{Trushkin}, {Nizhelskij}, {Tsybulev}  \&
  {Zhekanis}}{{Trushkin} et~al.}{2016b}]{Trushkin_2016ATel9416}
{Trushkin} S.~A.,  {Nizhelskij} N.~A.,  {Tsybulev} P.~G.,   {Zhekanis} G.~V.,
  2016b, The Astronomer's Telegram, \href
  {http://cdsads.u-strasbg.fr/abs/2016ATel.9416....1T} {9416}

\bibitem[\protect\citeauthoryear{{Trushkin}, {Nizhelskij}, {Tsybulev}  \&
  {Zhekanis}}{{Trushkin} et~al.}{2016c}]{Trushkin_2016ATel9501}
{Trushkin} S.~A.,  {Nizhelskij} N.~A.,  {Tsybulev} P.~G.,   {Zhekanis} G.~V.,
  2016c, The Astronomer's Telegram, \href
  {http://cdsads.u-strasbg.fr/abs/2016ATel.9501....1T} {9501}

\bibitem[\protect\citeauthoryear{{Tudose} et~al.,}{{Tudose}
  et~al.}{2007}]{Tudose_2007}
{Tudose} V.,  et~al., 2007, \mn@doi [\mnras]
  {10.1111/j.1745-3933.2006.00264.x}, \href
  {http://cdsads.u-strasbg.fr/abs/2007MNRAS.375L..11T} {375, L11}

\bibitem[\protect\citeauthoryear{{T{\"u}rler}}{{T{\"u}rler}}{2011}]{Turler_201%
1}
{T{\"u}rler} M.,  2011, \memsai, \href
  {http://cdsads.u-strasbg.fr/abs/2011MmSAI..82..104T} {82, 104}

\bibitem[\protect\citeauthoryear{{T{\"u}rler}, {Courvoisier}  \&
  {Paltani}}{{T{\"u}rler} et~al.}{1999}]{Turler_1999}
{T{\"u}rler} M.,  {Courvoisier} T.~J.-L.,   {Paltani} S.,  1999, \aap, \href
  {http://cdsads.u-strasbg.fr/abs/1999A%26A...349...45T} {349, 45}

\bibitem[\protect\citeauthoryear{{Uttley} \& {Klein-Wolt}}{{Uttley} \&
  {Klein-Wolt}}{2015}]{Uttley_2015}
{Uttley} P.,  {Klein-Wolt} M.,  2015, \mn@doi [\mnras] {10.1093/mnras/stv978},
  \href {http://cdsads.u-strasbg.fr/abs/2015MNRAS.451..475U} {451, 475}

\bibitem[\protect\citeauthoryear{{Waltman}, {Fiedler}, {Johnston}  \&
  {Ghigo}}{{Waltman} et~al.}{1994}]{Waltman_1994}
{Waltman} E.~B.,  {Fiedler} R.~L.,  {Johnston} K.~J.,   {Ghigo} F.~D.,  1994,
  \mn@doi [\aj] {10.1086/117056}, \href
  {http://cdsads.u-strasbg.fr/abs/1994AJ....108..179W} {108, 179}

\bibitem[\protect\citeauthoryear{{Waltman}, {Ghigo}, {Johnston}, {Foster},
  {Fiedler}  \& {Spencer}}{{Waltman} et~al.}{1995}]{Waltman_1995}
{Waltman} E.~B.,  {Ghigo} F.~D.,  {Johnston} K.~J.,  {Foster} R.~S.,  {Fiedler}
  R.~L.,   {Spencer} J.~H.,  1995, \mn@doi [\aj] {10.1086/117518}, \href
  {http://cdsads.u-strasbg.fr/abs/1995AJ....110..290W} {110, 290}

\bibitem[\protect\citeauthoryear{{Waltman}, {Foster}, {Pooley}, {Fender}  \&
  {Ghigo}}{{Waltman} et~al.}{1996}]{Waltman_1996}
{Waltman} E.~B.,  {Foster} R.~S.,  {Pooley} G.~G.,  {Fender} R.~P.,   {Ghigo}
  F.~D.,  1996, \mn@doi [\aj] {10.1086/118213}, \href
  {http://cdsads.u-strasbg.fr/abs/1996AJ....112.2690W} {112, 2690}

\bibitem[\protect\citeauthoryear{{Watanabe}, {Kitamoto}, {Miyamoto}, {Fielder},
  {Waltman}, {Johnston}  \& {Ghigo}}{{Watanabe} et~al.}{1994}]{Watanabe_1994}
{Watanabe} H.,  {Kitamoto} S.,  {Miyamoto} S.,  {Fielder} R.~L.,  {Waltman}
  E.~B.,  {Johnston} K.~J.,   {Ghigo} F.~D.,  1994, \mn@doi [\apj]
  {10.1086/174649}, \href {http://cdsads.u-strasbg.fr/abs/1994ApJ...433..350W}
  {433, 350}

\bibitem[\protect\citeauthoryear{{Zdziarski}, {Segreto}  \&
  {Pooley}}{{Zdziarski} et~al.}{2016}]{Zdziarski_2016}
{Zdziarski} A.~A.,  {Segreto} A.,   {Pooley} G.~G.,  2016, \mn@doi [\mnras]
  {10.1093/mnras/stv2647}, \href
  {http://cdsads.u-strasbg.fr/abs/2016MNRAS.456..775Z} {456, 775}

\bibitem[\protect\citeauthoryear{{van Kerkwijk}, {Geballe}, {King}, {van der
  Klis}  \& {van Paradijs}}{{van Kerkwijk} et~al.}{1996}]{vanKerkwijk_1996}
{van Kerkwijk} M.~H.,  {Geballe} T.~R.,  {King} D.~L.,  {van der Klis} M.,
  {van Paradijs} J.,  1996, \aap, \href
  {http://cdsads.u-strasbg.fr/abs/1996A%26A...314..521V} {314, 521}

\bibitem[\protect\citeauthoryear{{van der Laan}}{{van der
  Laan}}{1966}]{van-der-Laan_1966}
{van der Laan} H.,  1966, \mn@doi [\nat] {10.1038/2111131a0}, \href
  {http://cdsads.u-strasbg.fr/abs/1966Natur.211.1131V} {211, 1131}

\makeatother
\end{thebibliography}

%\begin{thebibliography}{99}
%\end{thebibliography}

%%%%%%%%%%%%%%%%%%%%%%%%%%%%%%%%%%%%%%%%%%%%%%%%%%

%%%%%%%%%%%%%%%%% APPENDICES %%%%%%%%%%%%%%%%%%%%%

%\appendix



%%%%%%%%%%%%%%%%%%%%%%%%%%%%%%%%%%%%%%%%%%%%%%%%%%

% Don't change these lines
\bsp	% typesetting comment
\label{lastpage}
\end{document}